\documentclass[aps,prd,superscriptaddress,nofootinbib,tighten,preprint]{revtex4}
\usepackage{color}
\usepackage{latexsym}
\usepackage{amssymb}
\usepackage{amsmath}
\usepackage{graphicx}
\usepackage{hyperref}
\usepackage{appendix}
\usepackage{grffile}
\begin{document}

\title{\boldmath Probing the sensitivity to leptonic $\delta_{CP}$ in presence of invisible decay of $\nu_3$ using 
atmospheric neutrinos}

\author{Lakshmi.S.Mohan}
\email{Lakshmi.Mohan@ncbj.gov.pl}
\affiliation{National Centre for Nuclear Research (NCBJ), Warsaw, Poland}
          
\date{\today} 
\bigskip

\begin{abstract}
One of the main neutrino oscillation parameters whose value has not been determined 
very precisely is the leptonic $\delta_{CP}$ phase. Since neutrinos have a tiny but finite mass 
they can undergo decay both visibly and invisibly. The effect of invisible decay of the third mass eigen 
state $\nu_3$ on the sensitivity to $\delta_{CP}$ is analysed here using 
atmospheric neutrino and anti-neutrino events. Effects of detector resolutions and systematic uncertainties 
are studied to identify the optimum resolutions and efficiencies required by 
a detector to obtain a significant sensitivity even in presence of decay. 
\end{abstract}

\maketitle
\section{Introduction}\label{intro}
    The value of the leptonic $\delta_{CP}$ phase is one of the most sought out unknowns in neutrino 
    oscillation physics. Several accelerator based long baseline (LBL) experiments are taking data 
    \cite{T2K-nature,nova} and are being planned \cite{dune,dedalus} to measure this quantity precisely. 
    Since neutrino oscillations have proven that neutrinos have a tiny but finite mass, there is a possibility 
    that they can decay. For Majorana neutrinos a possible decay mode allowed by Majoron model 
    \cite{majoron,majoron1,majoron2} is $\nu_i\to\nu_j+J$ or 
    $\nu_i\to\bar{\nu}_j+J$, where $\nu_j$ and $\bar{\nu}_j$ are lighter neutrino and anti--neutrino states 
    and J is mostly a singlet Majoron \cite{majoron1,majoron2}. This decay can be visible or invisible depending on 
    whether the final state contains an active neutrino or a sterile neutrino respectively. In this paper the effect of 
    invisible decay will be studied. The lifetimes of $\nu_1$ and $\nu_2$ are tightly constrained by solar neutrino 
    data \cite{Acker,Berezhiani,Berezhiani1,sc-sg-dm,band-sc-sg,joshipura,band-sc-sg-1,Berryman,Picoreti,SN1987A, Huang}.
    Detailed discussions on constraining neutrino lifetimes via cosmology and astrophysical experimental
    scales are performed in 
    \cite{beacom-astro,maltoni-flav-rat,Pagliaroli-non-rad,Bustamante,mohanty-sir,
    white,Hannestad1,Hannestad2,Escudero,Khlopov1,Khlopov2,Khlopov3,Pasquale,dror,Abdullahi,chacko}. Many of these 
    papers have considered the invisible neutrino decay scenario, especially the Majoron model. Analyses for 
    invisible neutrino decay for accelerator long and medium baseline and atmospheric neutrino experiments have been 
    carried out in \cite{gonzalez-garcia,gomes-gomes,dp,dp1,moment,anish-dune,sk-sc-sg,a3-pre,orca-decay} 
    to obtain the limits for $\nu_3$ decay. The limits from visible decay in accelerator and reactor neutrino experiments
    are discussed in \cite{coloma-vis,gago,suprabh,sosa}.
    
    In this study the decay of $\nu_3$ 
    into a light sterile neutrino state with which it does not mix \cite{sk-sc-sg,pakvasa-prl} is considered. So there 
    will be the dominant oscillations plus subdominant invisible decay. The study presented in this paper is mainly 
    concerned with neutrino energies and distances available for terrestrial experiments - the energies and baselines corresponding 
    to atmospheric neutrino experiments. The invisible decay will cause a depletion of 
    observable events in the detector. This decay is characterised by a parameter 
    $\alpha_3=m_3/\tau_3$, where $m_3$ and $\tau_3$ are the mass and rest frame life time of $\nu_3$ respectively. 
    It has been shown in \cite{gonzalez-garcia,gomes-gomes,dp,dp1,a3-pre,orca-decay} that 
    invisible decay will affect the measurement of other oscillation parameters, especially $\theta_{23}$. While the 
    effect of decay on the measurement of $\delta_{CP}$ with LBL experiments has been studied in \cite{dp}, this has not
    been studied in detail with atmospheric neutrinos. Atmospheric neutrinos offer a wide variety of baselines (from $\sim$15 km -- $\sim$ 13000 km)
    and energies ($\sim$0.1--30.0 GeV) of neutrinos. Studies on sensitivity to $\delta_{CP}$ using sub-GeV atmospheric
    neutrinos have been conducted in \cite{lowE-atmos-dcp,dune-atmos-delcp,wawu-dcp}. Low energy (sub-GeV) atmospheric 
    neutrinos are a good probe for the effect of this invisible decay of $\nu_3$ due to several reasons. 
    \begin{enumerate}
     \item Measurement of $\delta_{CP}$ unambiguous of neutrino mass hierarchy - The measurement of neutrino oscillation 
     parameters is affected by degeneracies. Presence of more parameters mean more degeneracies and ambiguities and 
     neutrino mass hierarchy has not been determined yet. At sub-GeV energies $\delta_{CP}$ can be measured unambiguous 
     of hierarchy \cite{lowE-atmos-dcp,dune-atmos-delcp,wawu-dcp}. This opens up the possibility of determining the effect of 
     other parameters like $\alpha_3$ on $\delta_{CP}$ measurement. 

     \item Effect of the invisible decay parameter $\alpha_3$ is more at lower (sub-GeV) energies \cite{a3-pre}. 
      Hence, in the absence of other degeneracies, its effects on $\delta_{CP}$ measurement will me more evident 
      at these energies.  
     
     \item Statistics - The flux of atmospheric neutrinos are large at sub-GeV energies\cite{honda,honda1,honda2}. Hence there will be more 
     number of events available for the study. 
     
    \end{enumerate}
 
    In this paper a study of how the presence of invisible decay of $\nu_3$ affects the measurement of $\delta_CP$ with 
    atmospheric neutrinos is conducted. The optimum detector configurations required to achieve a good sensitivity to 
    $\delta_{CP}$ even in the presence of $\alpha_3$ is also studied. The effect of invisible decay on the oscillation 
    probabilities and event spectra relevant for this study are discussed in Sections.\ref{posc-delcp-a3-eve-spec}.
    The process of event generation for different types of analyses are discussed in Section.\ref{eve-gen}. 
    Sensitivities to $\delta_{CP}$ in presence of decay for ideal and realistic cases in the absence and presence of 
    systematic uncertainties are discussed in Sections.\ref{ideal-chi2} and\ref{real-chi2} respectively. Summary and conclusions are given 
    in Section.\ref{summary}.

\section{Effect in visible decay on oscillation probabilities in matter}\label{posc-delcp-a3-eve-spec}
 A full 3-flavour oscillations + decay in matter scenario is considered here. 
 The mass eigen state $\nu_3$ decays invisibly via $\nu_3\rightarrow \nu_s+J$, where
 $J$ is a pseudo-scalar Majaron and $\nu_s$ is a sterile neutrino which does not mix with the 
 three active neutrinos. Hence the mixing matrix $U$ in vacuum will be:
\begin{equation}
 U = \begin{pmatrix}
c_{12}c_{13} & s_{12}c_{13} & s_{13}e^{-i\delta} \\
-c_{23}s_{12}-s_{23}s_{13}c_{12}e^{i\delta} & c_{23}c_{12}-s_{23}s_{13}s_{12}e^{i\delta} & s_{23}c_{13} \\
s_{23}s_{12}-c_{23}s_{13}c_{12}e^{i\delta} & -s_{23}c_{12}-c_{23}s_{13}s_{12}e^{i\delta} & c_{23}c_{13}
\end{pmatrix},  
\label{upmns}
\end{equation}
where $c_{ij}~=~\cos\theta_{ij}$, $s_{ij}~=~\sin\theta_{ij}$; $\theta_{ij}$ are the mixing angles and 
$\delta$ is the CP violating phase. 

For true normal hierarchy, $m_s<m_1<m_2<m_3$, where $m_s$ is the mass of $\nu_s$ and $m_i$ are the mass of 
$\nu_i$, i=1,2,3. In the presence of Earth matter, the three-flavor evolution equation will be:
 \begin{equation}
  i\frac{d\tilde{\nu}}{dt} = \frac{1}{2E}\left[U\mathbb{M}^2U^\dagger + \mathbb{A}_{CC}\right]\tilde{\nu},
  \label{nuevematter}
  \end{equation}
  \begin{equation}
  \mathbb{M}^2~=~
  \begin{pmatrix}
  0 & 0 & 0 \\ 
  0 & \Delta{m^2_{21}} & 0 \\ 
  0 & 0 & \Delta{m^2_{31}}-i\alpha_3 
  \end{pmatrix}
  \,,~~{\rm and}~~
    \mathbb{A}_{CC}~=~
  \begin{pmatrix}
  A_{cc} & 0 & 0 \\ 
  0 & 0 & 0 \\ 
  0 & 0 & 0 
  \end{pmatrix}
  ,
 \end{equation}
where $E$ is the neutrino energy, $\alpha_3=m_3/\tau_3$ is the decay constant in units of eV$^2$,
$m_3$ is the mass of $\nu_3$ and $\tau_3$ its rest frame life time and $A_{cc}$ is the matter potential. 
\begin{equation}
A_{cc}=2\sqrt{2}G_Fn_eE=7.63\times10^{-5}\hbox{eV}^2~\rho(\hbox{gm/cc})~E(\hbox{GeV})
\label{matt-pot}
\end{equation}
where, $G_F$ is the Fermi constant and $n_e$ is the electron number density in matter and $\rho$ is the matter 
density. For anti-neutrinos, both the sign of $A_{cc}$ and the phase $\delta$ in Eq.~(\ref{nuevematter}) are
reversed. Since the term $\alpha_3$ appears in the propagation
equation along with $\Delta{m^2_{31}}$, they should have the same unit. The conversion factor 
to make $\alpha_3$ and $\Delta{m^2_{31}}$ have the same units (i.e $eV^2$) is 
$1~ \hbox{eV/s} = 6.58\times10^{-16}~ \hbox{eV}^2$.

\begin{figure}[htp]
\includegraphics[width=0.45\textwidth,height=0.45\textwidth]{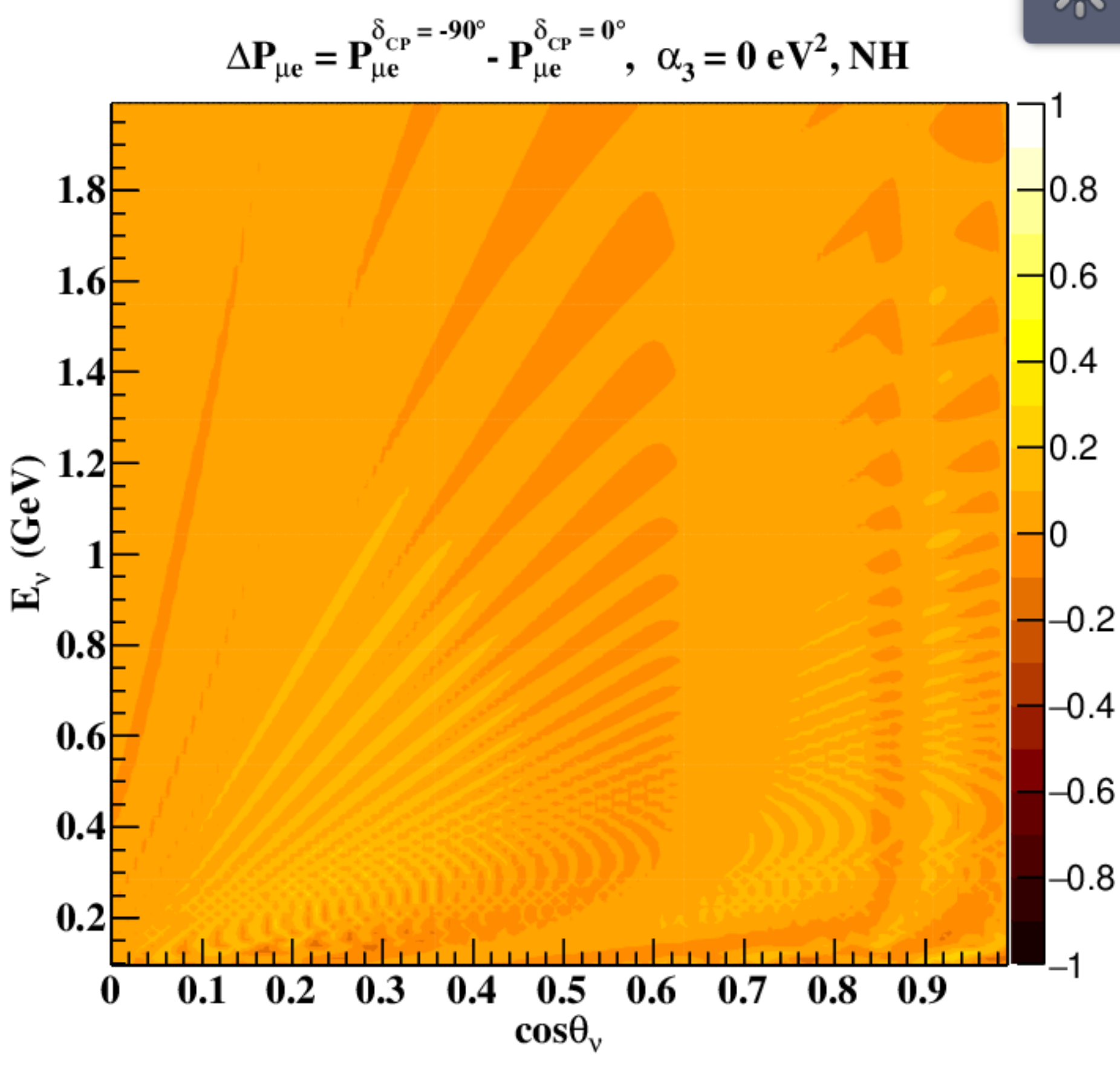}
\includegraphics[width=0.45\textwidth,height=0.45\textwidth]{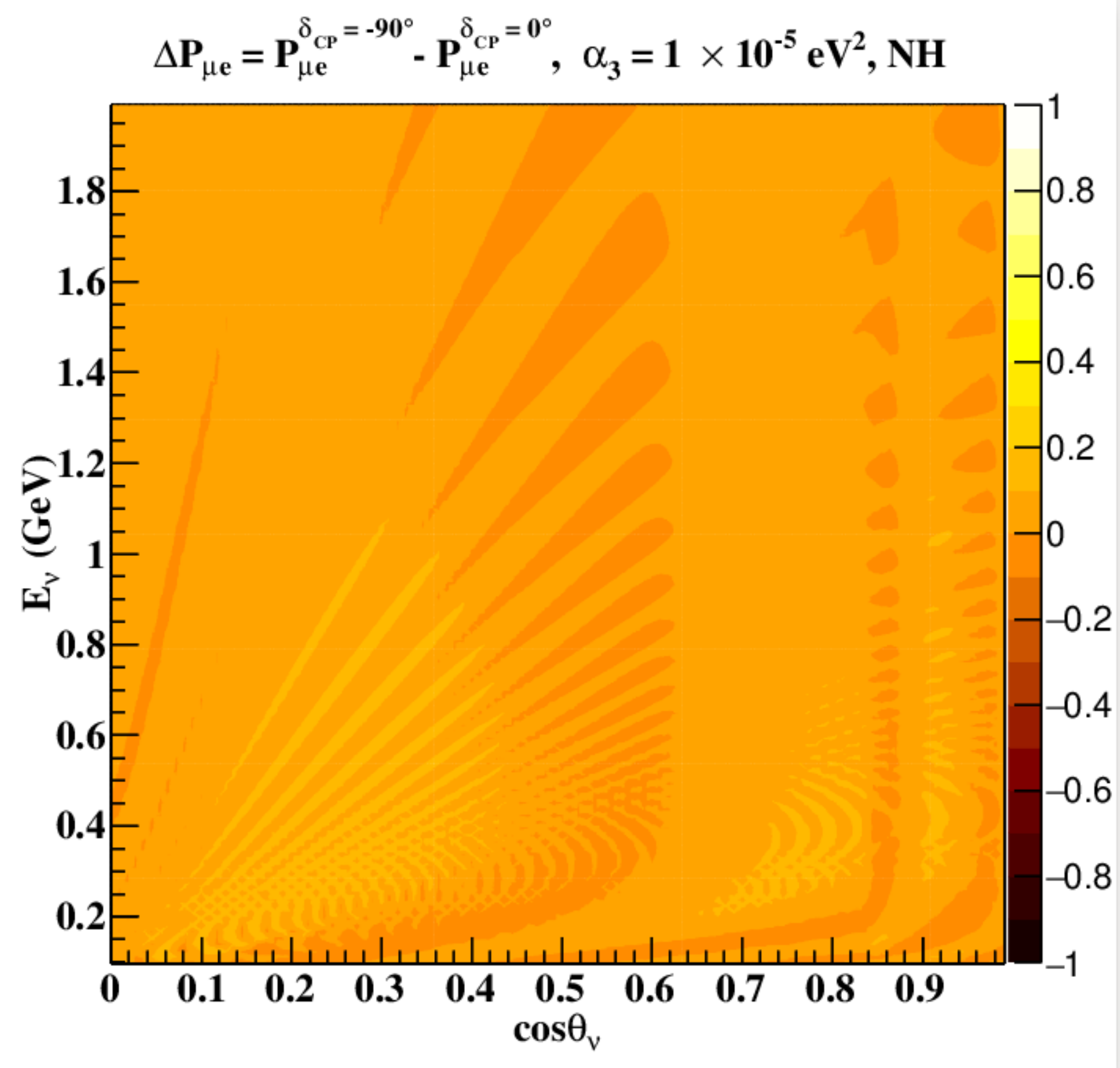}

\includegraphics[width=0.45\textwidth,height=0.45\textwidth]{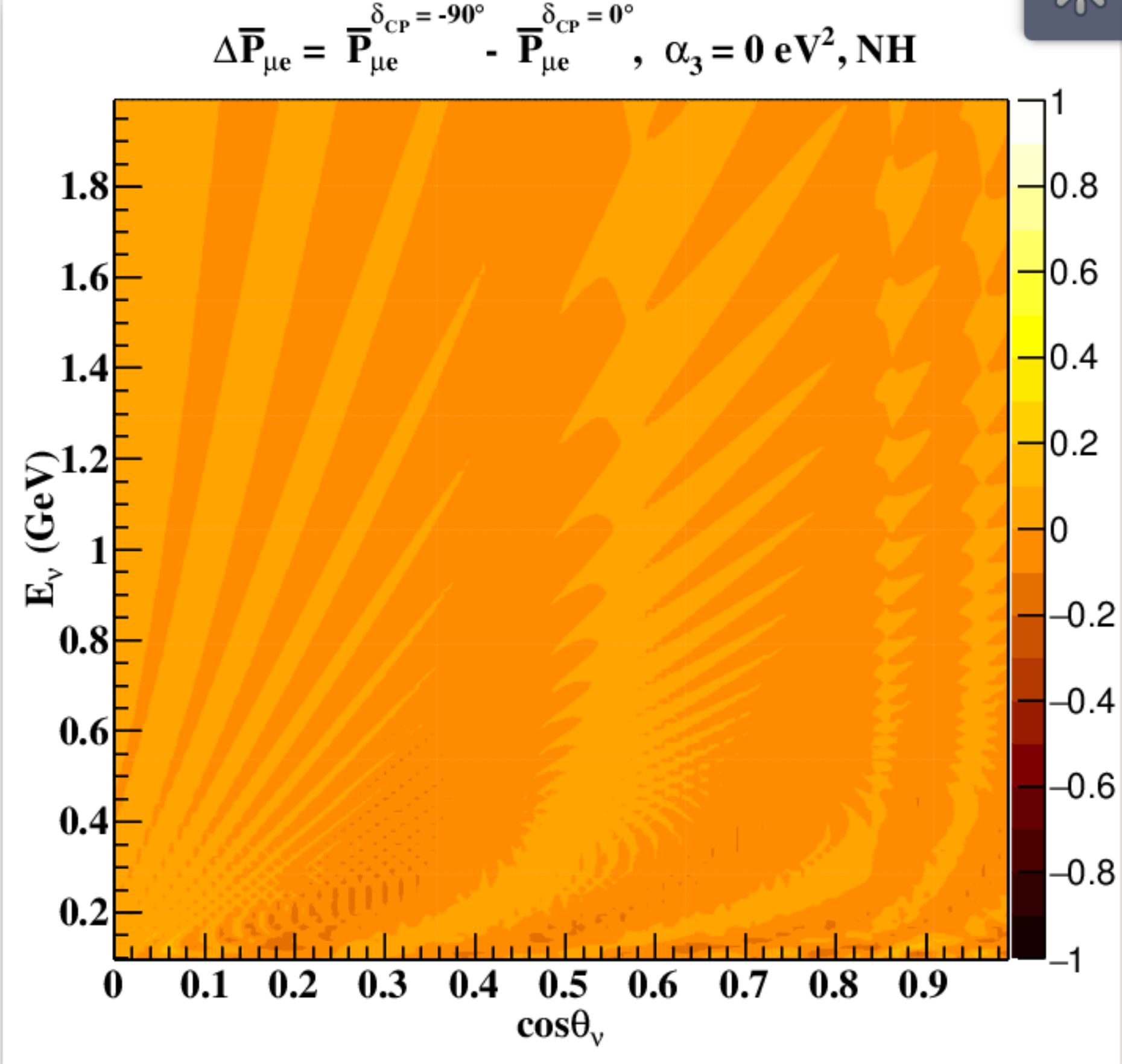}
\includegraphics[width=0.45\textwidth,height=0.45\textwidth]{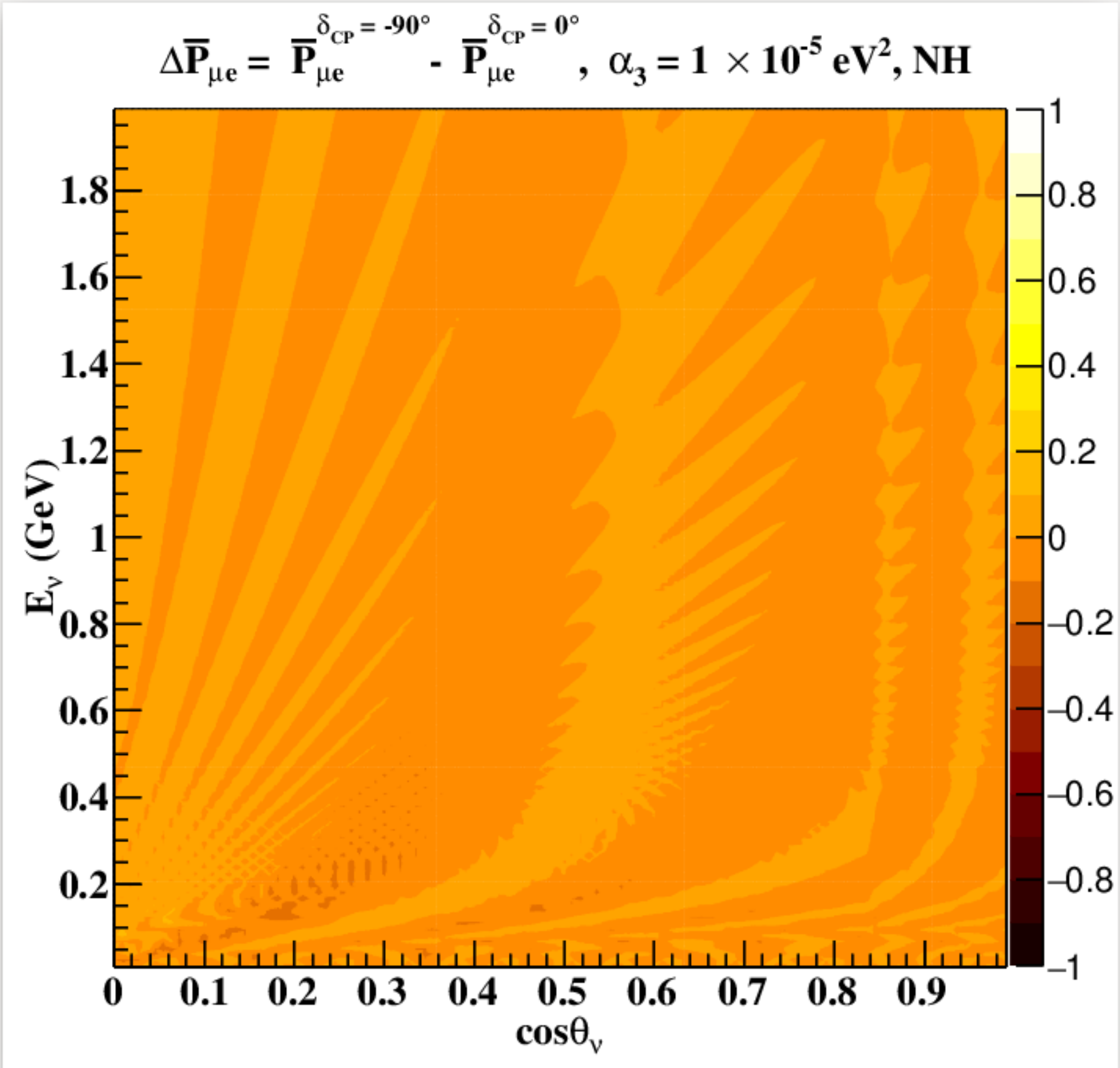}
\caption{Top panels - $\Delta{P}_{\mu{e}}$ (left) for $\alpha_3=0$ eV$^2$ (no decay) and (right) 
$\alpha_3=1\times10^{-5}$ eV$^2$ in the neutrino energy range 0.1--2.0 GeV for true normal hierarchy.
Bottom panels are for $\Delta\bar{P}_{\mu{e}}$.}
\label{delPmue-delcp-a3}
\end{figure}

\begin{figure}[htp]
\includegraphics[width=0.45\textwidth,height=0.45\textwidth]{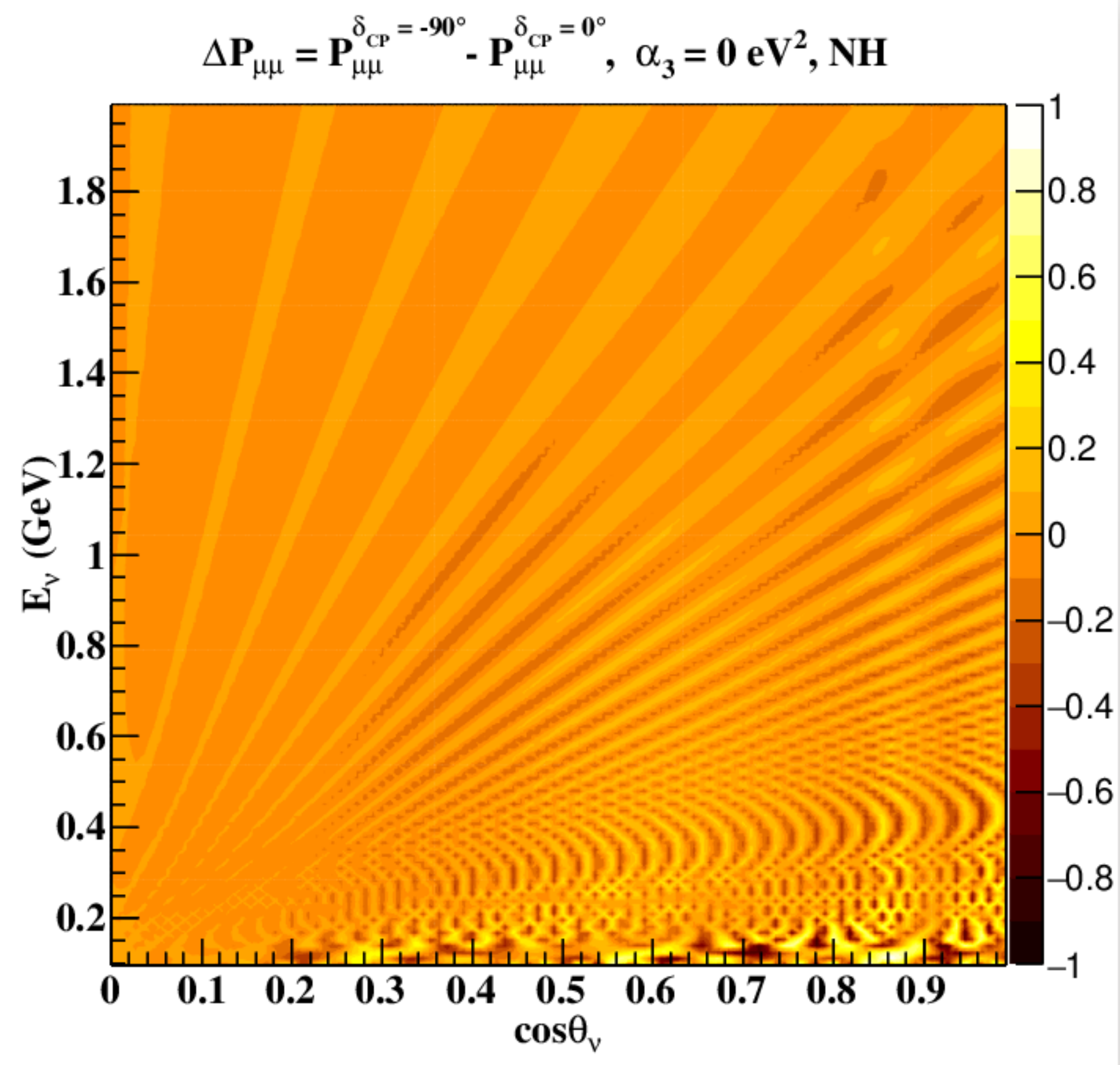}
\includegraphics[width=0.45\textwidth,height=0.45\textwidth]{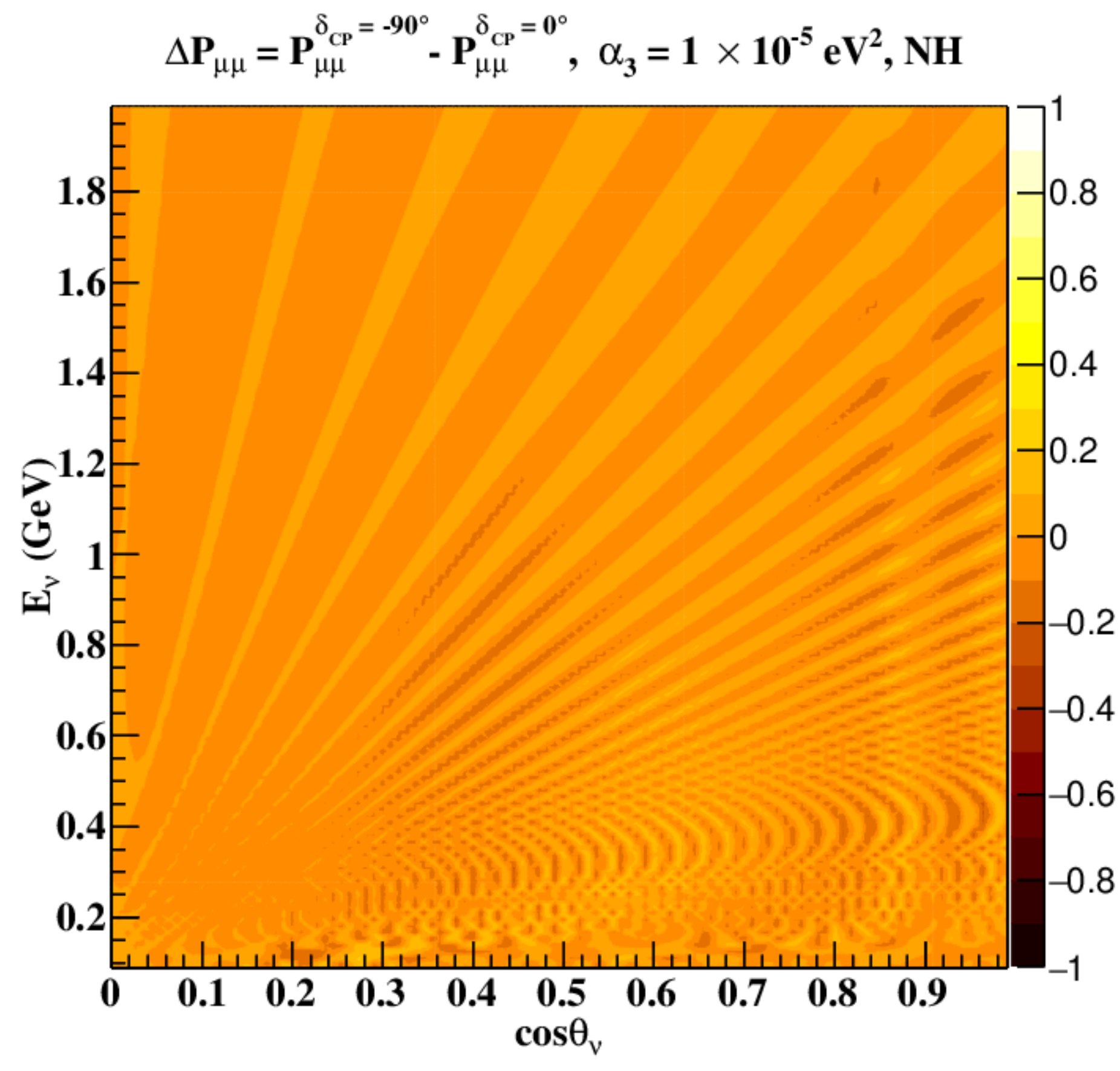}

\includegraphics[width=0.45\textwidth,height=0.45\textwidth]{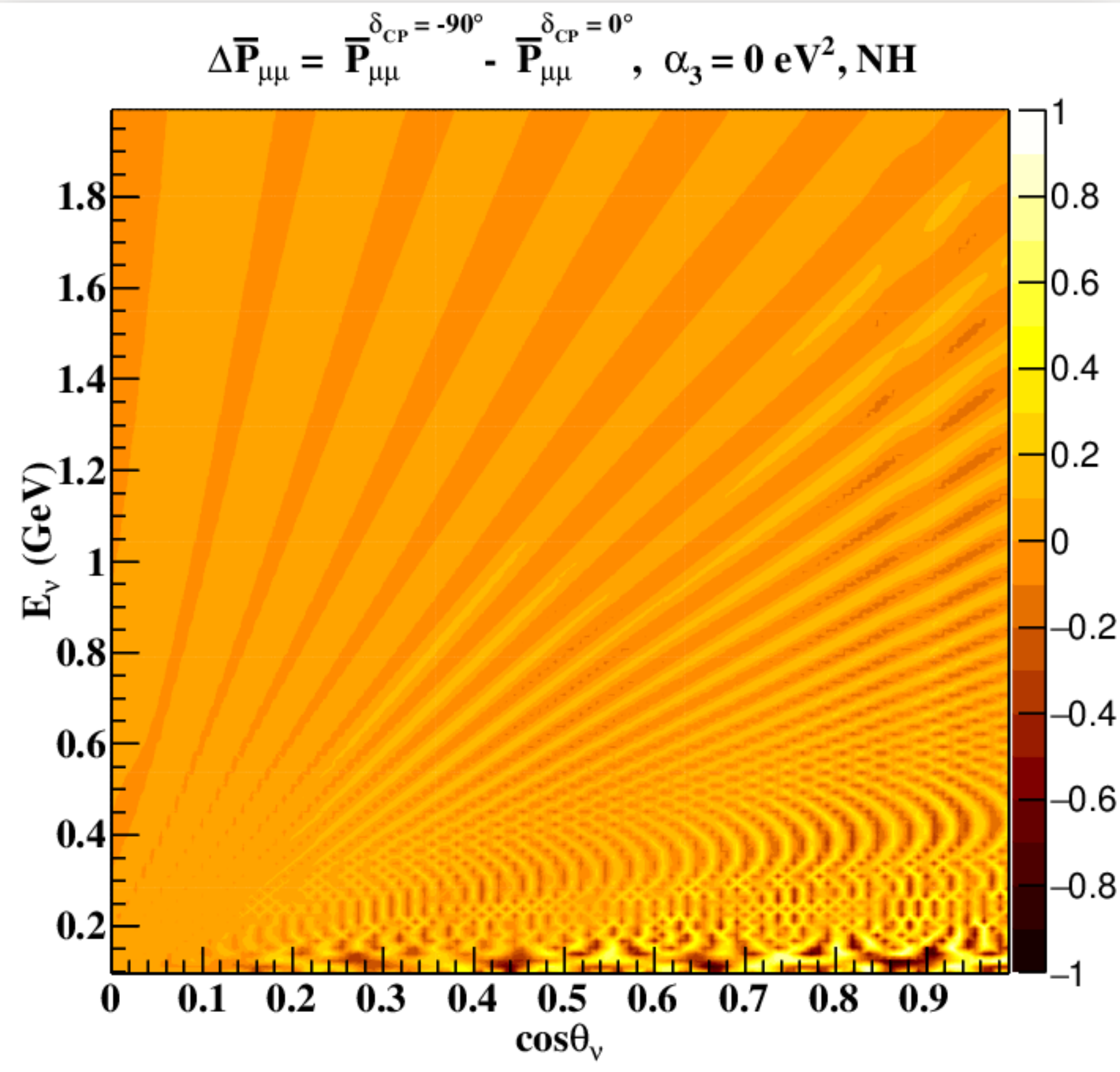}
\includegraphics[width=0.45\textwidth,height=0.45\textwidth]{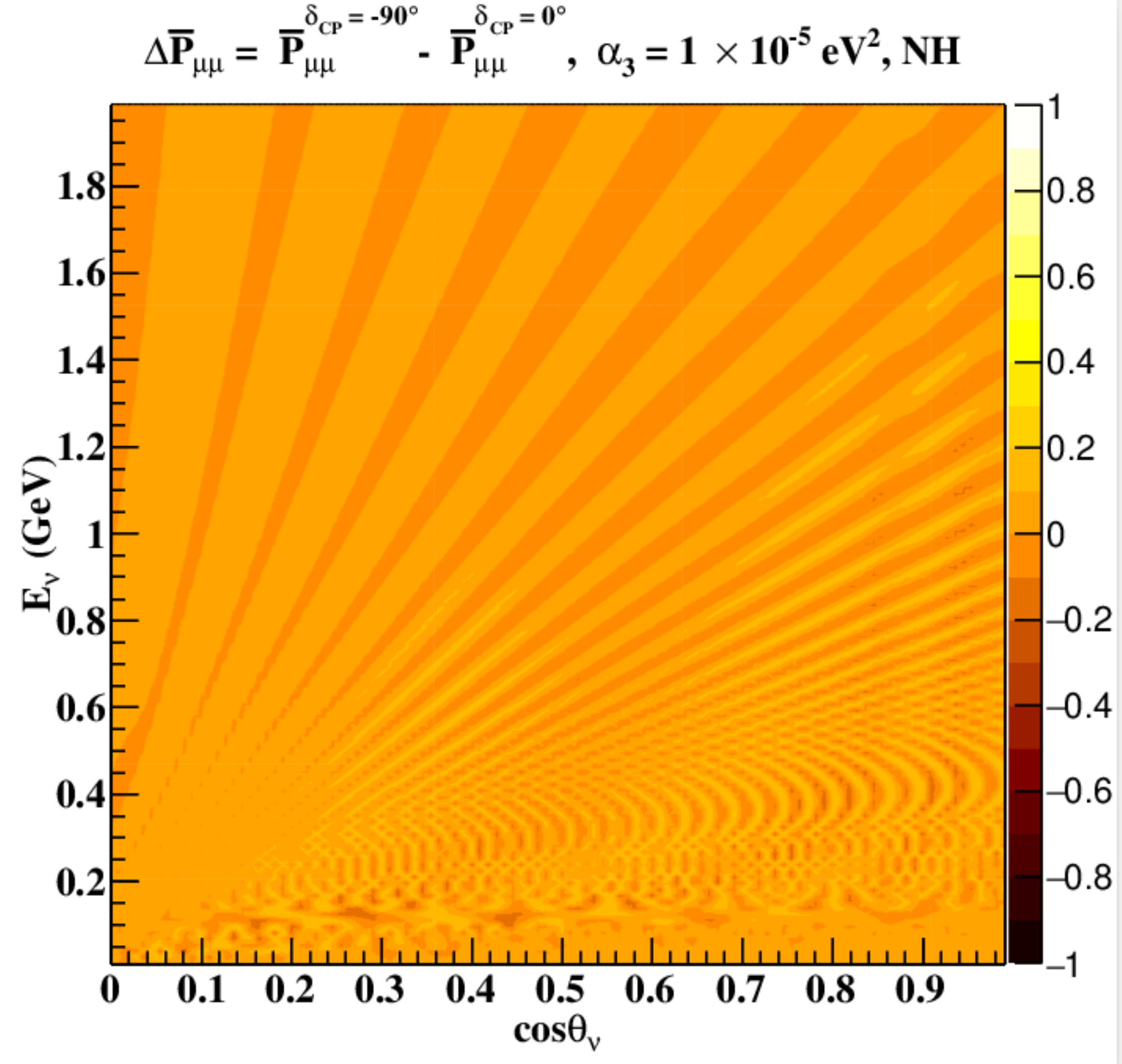}
\caption{Top panels - $\Delta{P}_{\mu\mu}$ (left) for $\alpha_3=0$ eV$^2$ (no decay) and (right) 
$\alpha_3=1\times10^{-5}$ eV$^2$ in the neutrino energy range 0.1--2.0 GeV for true normal hierarchy.
Bottom panels are for $\Delta\bar{P}_{\mu\mu}$.}
\label{delPmumu-delcp-a3}
\end{figure}

     Transition probabilities, especially are mainly responsible for the sensitivity to $\delta_{CP}$ 
     \cite{trans-prob,trans-prob-1,trans-prob-2}.
     Since $\Phi_{\nu_\mu}/\Phi_{\nu~e}(\Phi_{\overline{\nu_\mu}}/\Phi_{\overline{\nu_e}})\approx2:1$, the contribution
     to $\delta_{CP}$ sensitivity from $\nu_\mu\to\nu_e$ ($\overline{\nu_\mu}\to\overline{\nu_e}$) events will be more.
     The sensitivity to $\alpha_3$ is more for $\nu_\mu\to\nu_\mu$ and $\overline{\nu_\mu}\to\overline{\nu_\mu}$ events, 
     though their sensitivity to $\delta_{CP}$ is very low compared to the $\nu_e$ like events. Hence not to leave out 
     any contribution from any channel all 8 channels -
     $\nu_{e\beta},\overline{\nu_{e\beta}},\nu_{\mu\beta},\overline{\nu_{\mu\beta}}$, where $\beta=e,\mu$ are studied for 
     this analysis. The difference between the 3--flavour oscillation probabilities in matter with 
     $\delta_{CP}=-90^\circ$ and $\delta_{CP}=0^\circ$ is shown as oscillograms in Figs.~\ref{delPmue-delcp-a3} and 
     \ref{delPmumu-delcp-a3}. The oscillograms for the lower energy range 0.1--2.0 GeV are shown. 
     The central values of the oscillation parameters used for the analysis are given in Table.~\ref{osc-par-3sig}.
     \begin{table}[htp] \centering \begin{tabular}{|c|c|c|} 
   \hline
   Parameter & Input value & Marginalization range \\ 
   \hline 
   $\theta_{13}$ & 8.63$^\circ$ & Not marginalised \\ 
   $\sin^{2}\theta_{23}$ & 0.762889 & [0.42, 0.74] \\ 
   $\Delta{m^2_{eff}}$ & $2.56\times10^{-3}~{\rm eV}^2$ & [2.43, 2.79]$\times10^{-3}~{\rm eV}^2$\\ 
   $\sin^{2}\theta_{12}$ & 0.31 & Not marginalised \\ 
   $\Delta{m^{2}_{21}}$ & $7.39\times10^{-5}~{\rm eV}^2$ & Not marginalised \\ 
   $\delta_{CP}$ & $\pm90^\circ$ & Not marginalised \\ 
   \hline
\end{tabular} 
\caption{\small Central values of oscillation parameters and their 3$\sigma$ ranges used to 
generate oscillation probabilities in matter in presence of invisible decay of $\nu_3$. 
Central values of $\sin^2\theta_{23}$ and $\Delta{m^2}_{eff}$ are taken according to \cite{nufit-org}
and their 3$\sigma$ values from \cite{dp1}. The other oscillation 
parameters are taken from \cite{nufit-org}. Three different values of $\alpha_3$ are used,
$0$ (no decay), $4.36\times10^{3}$ eV$^2$ (90\% C.L from \cite{a3-pre}) and $1\times10^{-5}$ eV$^2$
(which is the approximate order of magnitude of limits on $\alpha_3$ obtained from LBL experiments like 
DUNE \cite{dp}). The 3$\sigma$ limit on $\alpha_3$ is taken as [0, $4.387\times^{-4}$] eV$^2$.
For the analysis, $\Delta m^2_{31} = \Delta m^2_{\rm eff}+\Delta~m_{21}^2\left(\cos^2\theta_{12}-\cos\delta_{CP}\sin\theta_{13}
\sin2\theta_{12}\tan\theta_{23}\right)~; \Delta m^2_{32}  =  \Delta~m^2_{31}-\Delta m^2_{21}$, for normal hierarchy 
with $\Delta{m}_{\rm~eff}^2 > 0$. $\Delta m^2_{31}\leftrightarrow -\Delta m^2_{32}$ for inverted hierarchy
when $\Delta{m}_{\rm eff}^2 < 0$.}
\label{osc-par-3sig} 
\end{table}

It can be seen from the Fig.~\ref{delPmue-delcp-a3} that $\alpha_3$ has no significant effect on $P_{\mu{e}}$
and $\bar{P}_{\mu{e}}$. However, the minor changes in the several bins with the increase of $\alpha_3$ value can add upto 
a small contribution to the $\delta_{CP}$ sensitivity. As seen in Fig.~\ref{delPmumu-delcp-a3}, $\alpha_3$ does affect 
$P_{\mu\mu}$ and $\bar{P}_{\mu\mu}$ at very low energies, especially below $E_\nu$ = 0.4 GeV. But the overall contribution 
of $P_{\mu\mu}$ and $\bar{P}_{\mu\mu}$ to $\delta_{CP}$ sensitivity is smaller compared to the electron like events. 
Atmopsheric neutrino flux is lesser compared to accelerator based long base line experiments and neutrino physics 
experiments are low counting experiments. So even the smallest contribution to the sensitivity to a parameter cannot be 
neglected. Because of this, the study is done for events in the neutrino energy range 0.1--30 GeV. Normal hierarchy 
is assumed to be the true hierarchy. At low energies $\delta_{CP}$ measurement will be independent of hierarchy 
\cite{lowE-atmos-dcp}, but at higher energies (2.0-30 GeV) the effect of hierarchy will be present, hence the hierarchy is 
assumed to be known, for this study. 
  
     For low energies 0.1--1.0 GeV, both $P_{\mu{e}}$ and $\overline{P}_{\mu{e}}$ are affected by invisible decay. In 
     the higher energy region, decay affects $P_{\mu e}$ in the resonance region. The effect of $\alpha_3$ on 
     $\overline{P}_{\mu{e}}$ is not as much as for the neutrino case. From this figure we can 
     see that the measurement of $\delta_{CP}$ will be affected by the presence of $\alpha_3$. The variation in
     sensitivity will depend on the value of $\alpha_3$, a lesser sensitivity is expected for larger $\alpha_3$ 
     from electron like events. 
     
\subsection{Effect of invisible decay of $\nu_3$ on the oscillated event spectra}
\begin{figure}
\includegraphics[width=0.45\textwidth,height=0.45\textwidth]{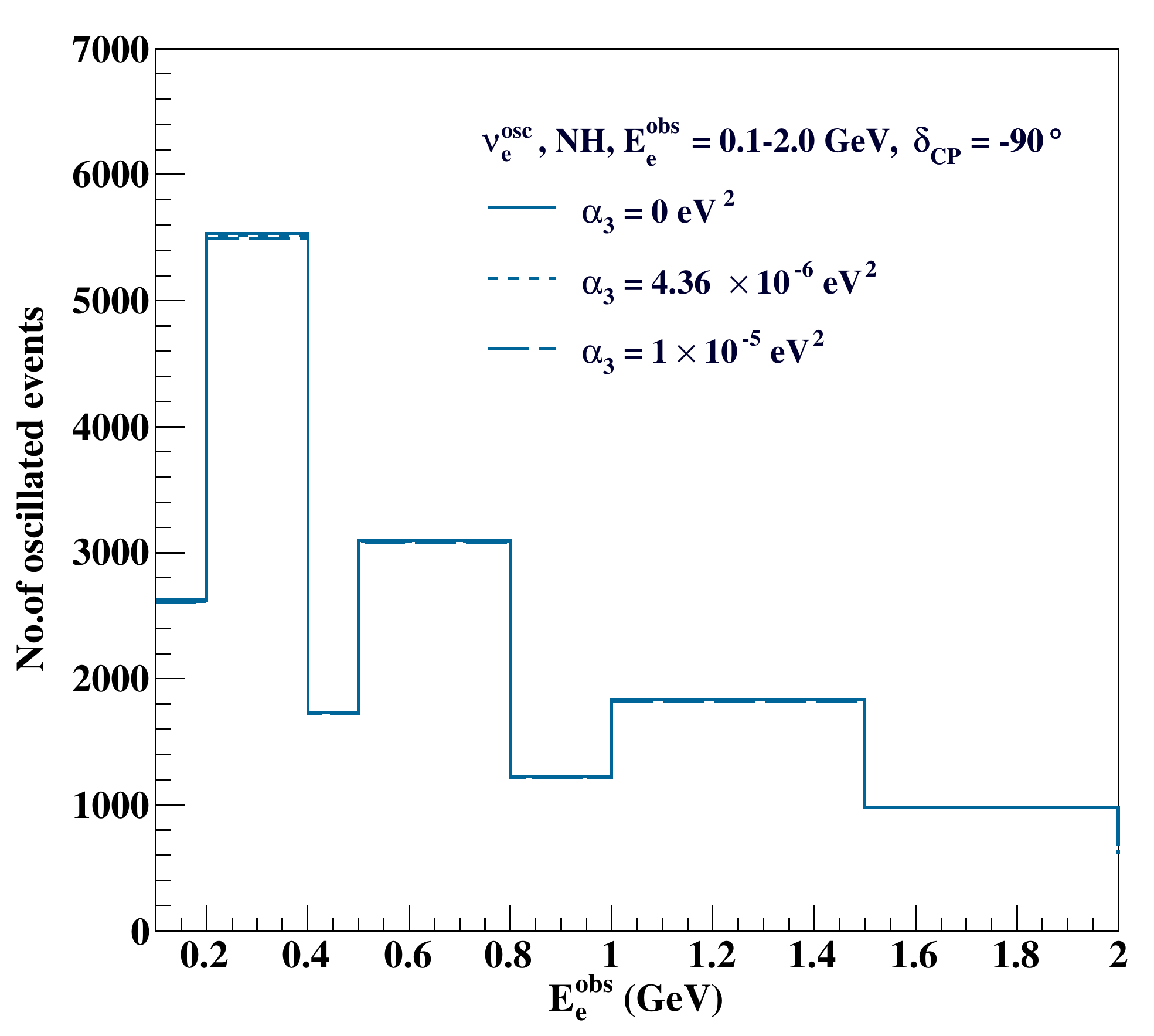}
\includegraphics[width=0.45\textwidth,height=0.45\textwidth]{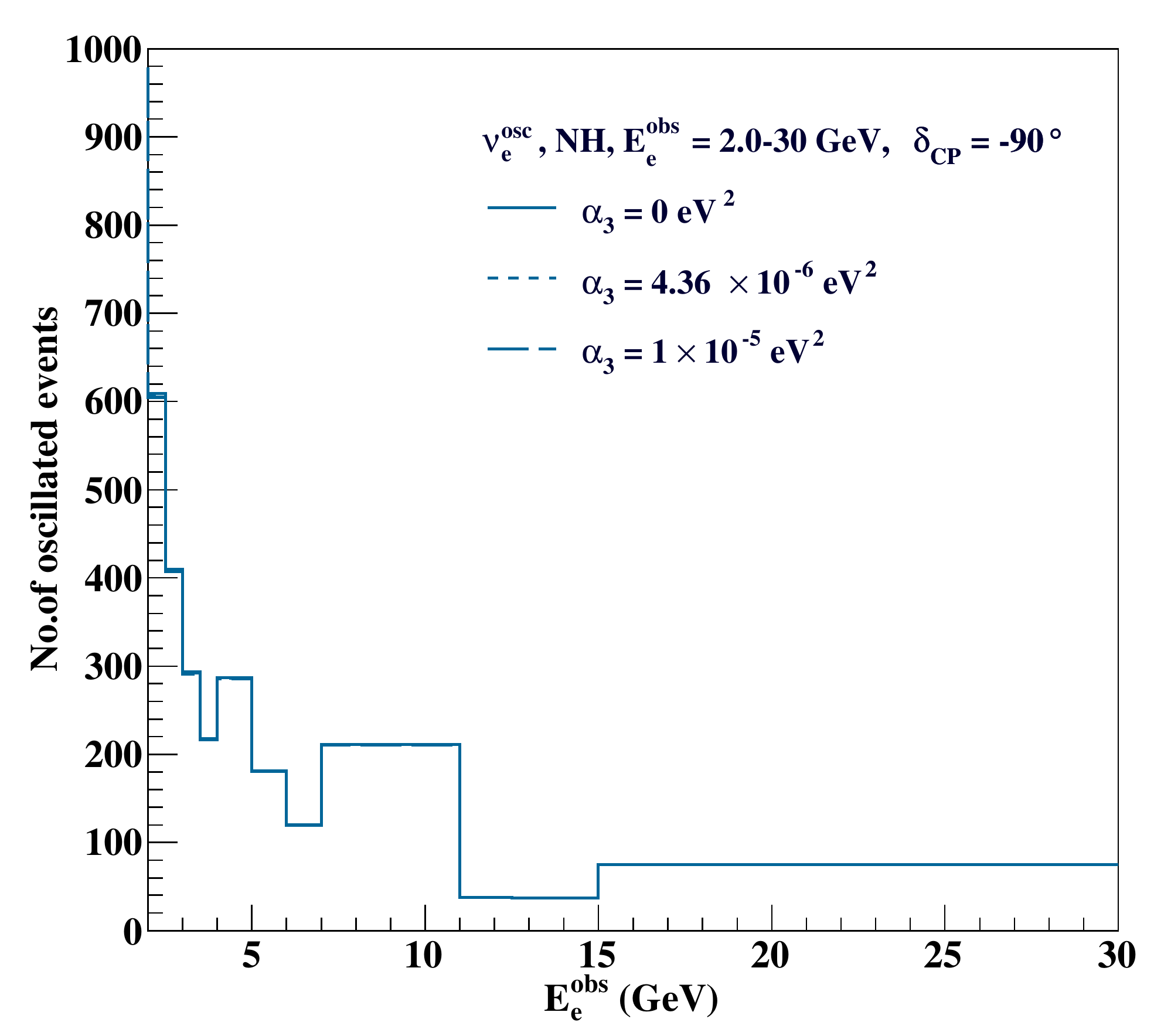}

\includegraphics[width=0.45\textwidth,height=0.45\textwidth]{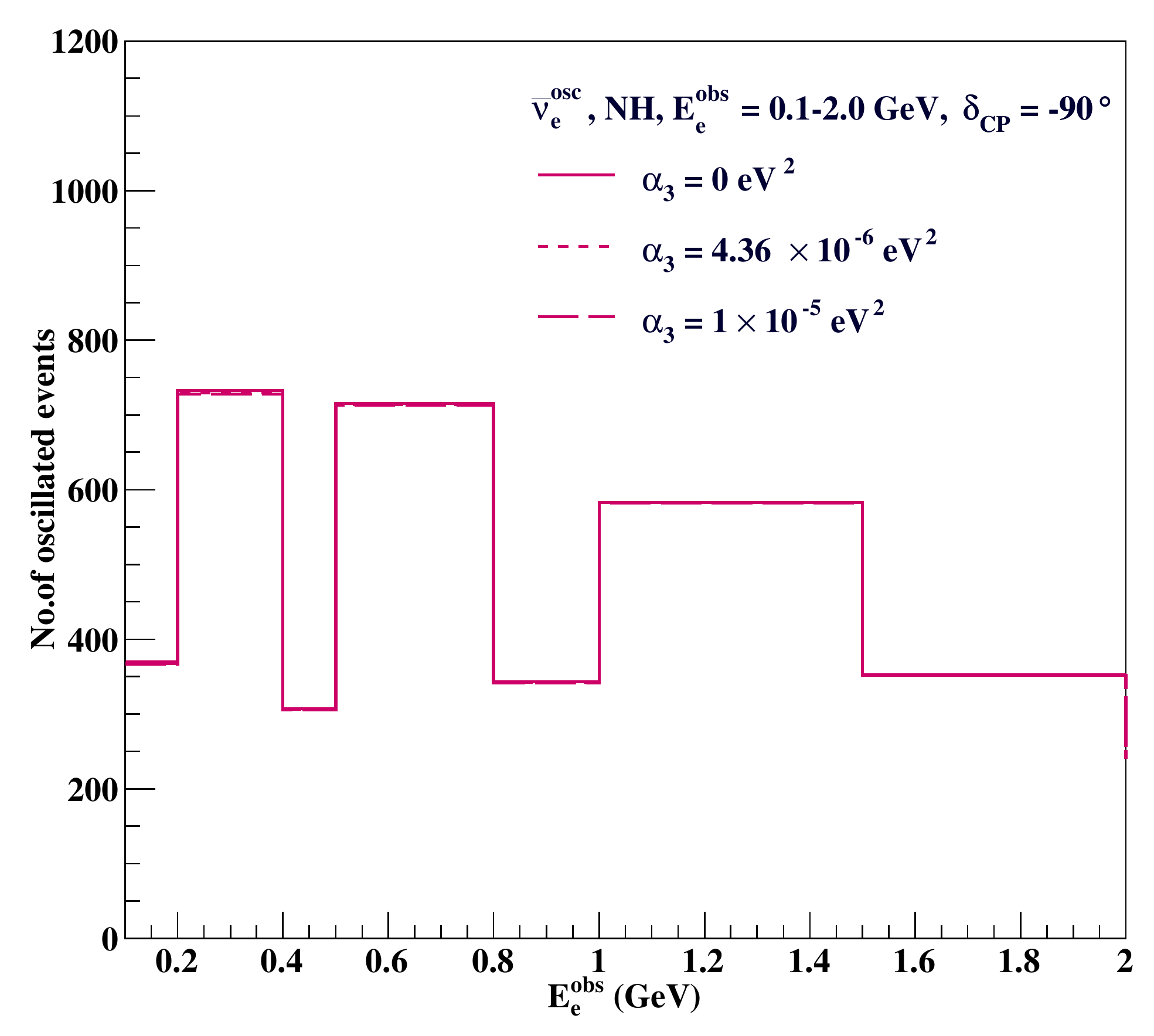}
\includegraphics[width=0.45\textwidth,height=0.45\textwidth]{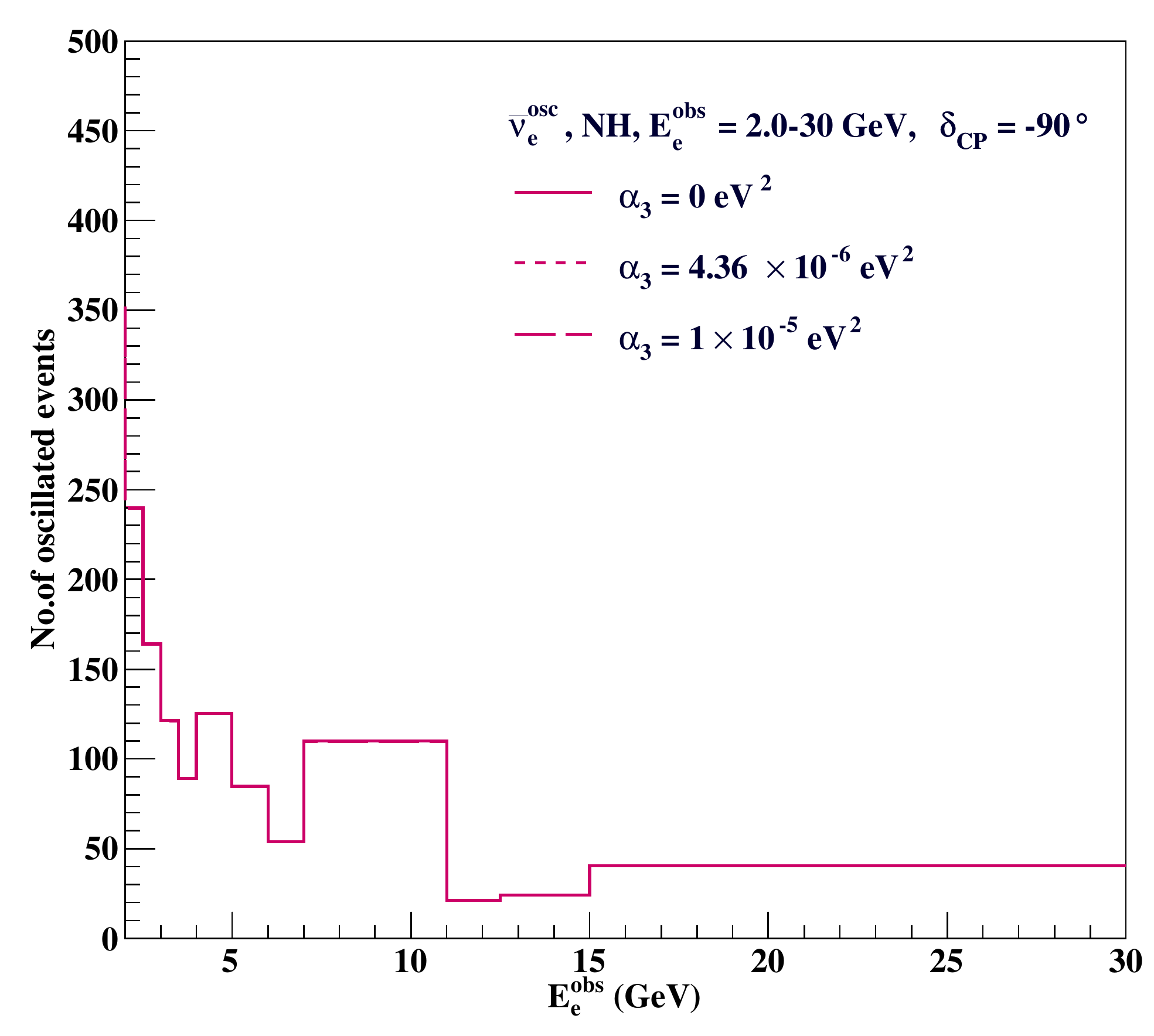}
\caption{Comparison of oscillated event spectra for different values of $\alpha_3$ for $\delta_{CP}=-90^\circ$.
Top panels are for $\nu_e$ and bottom panels are for $\overline{\nu}_e$ events. $\alpha_3$ has no effect on 
the oscillated spectra in both the energy ranges.}
\label{osc-spec-e}
\end{figure}

\begin{figure}
\includegraphics[width=0.45\textwidth,height=0.45\textwidth]{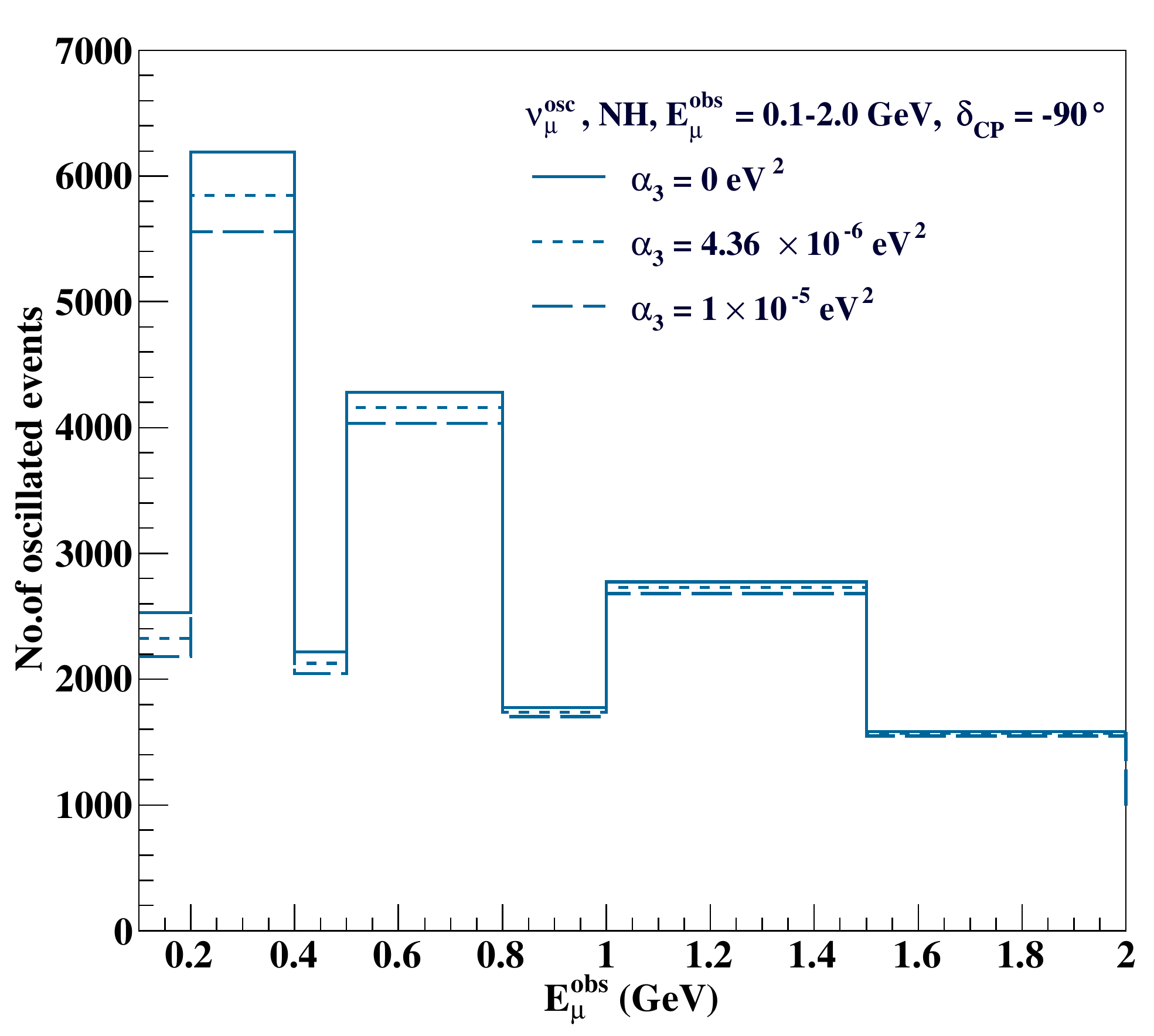}
\includegraphics[width=0.45\textwidth,height=0.45\textwidth]{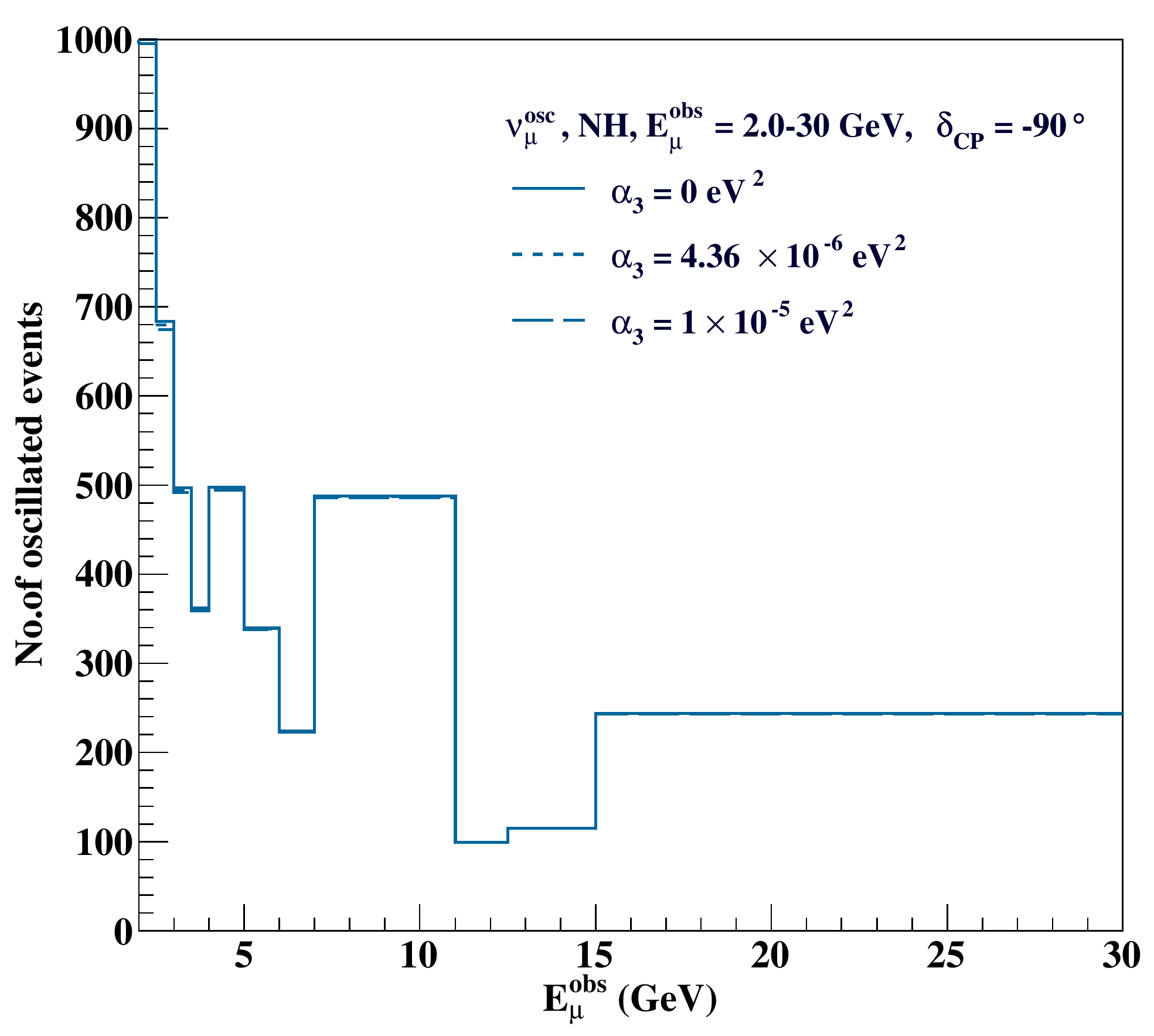}

\includegraphics[width=0.45\textwidth,height=0.45\textwidth]{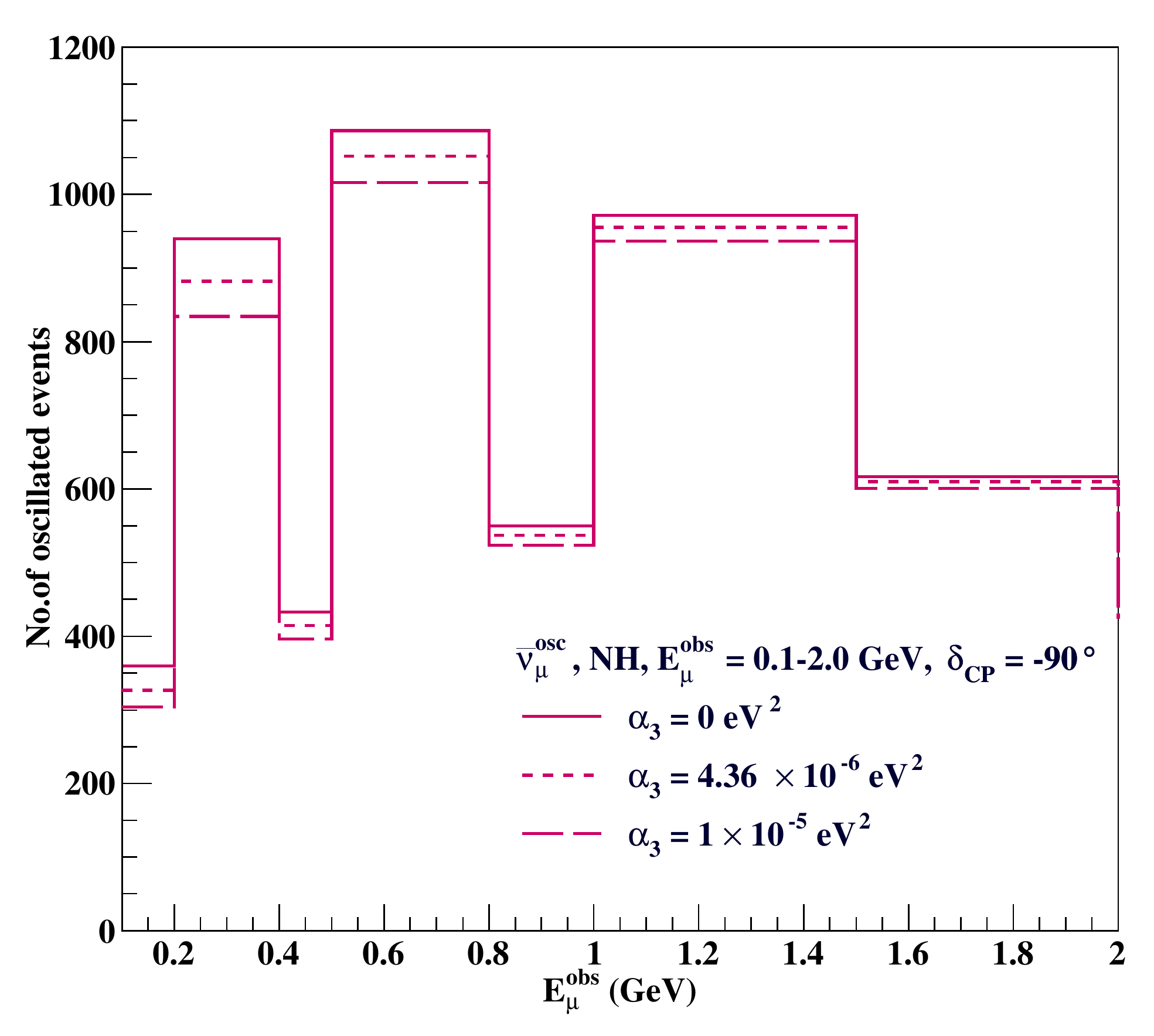}
\includegraphics[width=0.45\textwidth,height=0.45\textwidth]{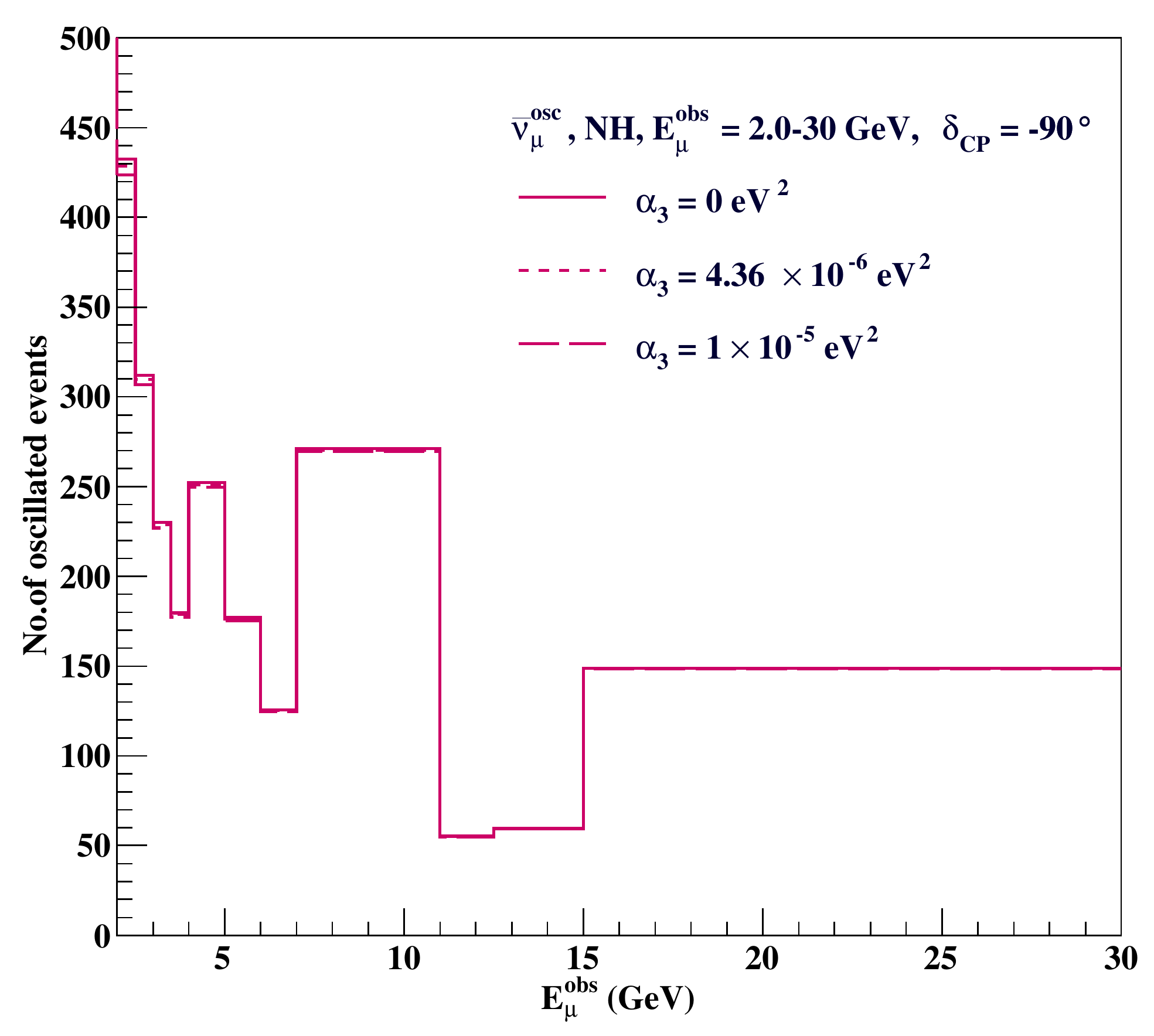}
\caption{Comparison of oscillated event spectra for different values of $\alpha_3$ for $\delta_{CP}=-90^\circ$.
Top panels are for $\nu_\mu$ and bottom panels are for $\overline{\nu}_\mu$ events. It is clear that $\alpha_3$
affects the $\nu_\mu$ and $\overline{\nu}_\mu$ spectra at low energies, which means that the sensitivity to 
$\delta_{CP}$ from these events, if any at all will be affected by $\alpha_3$.}
\label{osc-spec-mu}
\end{figure}

 The effect of the decay parameter $\alpha_3$ on the oscillated event spectra for different values of $\alpha_3$ and 
 $\delta_{CP}=-90^\circ$ is shown in Fig.~\ref{osc-spec-e},\ref{osc-spec-mu}. The effect of $\alpha_3$ in the 
 lower (0.1--2.0 GeV) and higher (2.0--30.0 GeV) energy regions can be separated. For $\nu_e$ and $\overline{\nu}_e$
 events, $\alpha_3$ does not have any significant effect, both at lower and higher energies. But the oscillated event 
 spectra get suppressed with increasing $\alpha_3$ values for both $\nu_\mu$ and $\overline{\nu}_\mu$ events. 
 Thus, the sensitivity to $\delta_{CP}$ from $\nu_e$ and $\overline{\nu}_e$ events will not get affected much by 
 $\alpha_3$, but the minor sensitivity from $\nu_\mu$ and $\overline{\nu}_\mu$ events will be. Electron like events 
 are more sensitive to $\delta_{CP}$ than muon like events. It can also be seen that the muon like events are more 
 sensitive to $\alpha_3$ especially at low energies. This shows that at low energies (0.1--2.0 GeV), $\nu_e$ and 
 $\overline{\nu}_e$ events are well suited to probe $\delta_{CP}$ while low energy $\nu_\mu$ and $\overline{\nu}_\mu$ 
 events probe $\alpha_3$ better. At higher energies the effects are not much, but the small contributions from all bins 
 can add up together. 
 
\clearpage
\section{Event generation and $\chi^2$ analysis}\label{eve-gen} 
  
  Simulated charged current (CC) $\nu_\mu,\overline{\nu}_\mu,\nu_e$ and
$\overline{\nu}_e$ events on an isoscalar target are used for this study. 
For atmospheric neutrinos, $\nu_e\rightarrow\nu_e$ survived events along with 
those from $\nu_\mu\rightarrow\nu_e$ transitions contribute to the (CC) $\nu_e$ 
event spectrum in the detector:
\begin{equation}
{\cal {N}}^e =t\times{n_d}\times \int d\sigma_{\nu_e}\times
   \left[P^m_{ee}\frac{d^2\Phi_e}{dE_\nu~d\cos\theta_\nu}+
   P^m_{\mu e}\frac{d^2\Phi_\mu} {dE_\nu~d\cos\theta_\nu}\right].
\label{toteve}
\end{equation}
Here $P^m_{ee}$ and $P^m_{\mu e}$ are the oscillation probabilities in matter in presence 
of decay. Here $t$ is the exposure/run time, $n_d$ is the number of targets available 
for interaction in the detector, $d\sigma_{\nu_e}$ is the neutrino interaction cross
section which is differential in final state charged lepton energy ($E_e$) and/or 
direction $\cos\theta_e$, and $d\Phi_{\nu_\mu}$ ($d\Phi_{\nu_e}$) is the 
$\nu_\mu$ ($\nu_e$) flux. Similarly for $\overline{\nu}_e$ and $\nu_\mu$ and 
$\overline{\nu}_\mu$ events also. Hereafter the charged current electron (muon) like events will be 
referred to as CCE (CCMU).

Sensitivity to $\delta_{CP}$ is studied for \emph{idealistic} and \emph{realistic} scenarios.
The difference between these scenarios is given in Table.~\ref{ideal-real}. 

\begin{table}[htp]
\begin{tabular}{|c|c|}
\hline
Idealistic & Realistic   \\
\hline
Perfect energy and direction resolution & Realistic energy \\
 for the final state particles (nores) & for final state particles (wres) \\
\hline
Complete separation of $\nu_e$ ($\bar\nu_e$) like & $\nu_e$ and $\nu_\mu$ like events can be separated from \\
from $\nu_\mu$ ($\bar\nu_\mu$) like events & each other \\
$\nu$ and $\bar{\nu}$ can be separated from each other (wcid) & No separation between $\nu$ and $\bar{\nu}$ (nocid) \\
\hline
Events binned in $(E^{obs}_l,\cos\theta^{obs}_{l},E^{obs}_{had'})$ (3D) & Events binned in $(E^{obs}_l,\cos\theta^{obs}_{l})$ (2D)\\
\hline
No fluctuations & With fluctuations \\
\hline 
\end{tabular}
\caption{Criteria for idealistic and realistic cases for sensitivity studies}
\label{ideal-real}
\end{table}

Unoscillated charged current (CC) events for an exposure of 100 years in a 
50 kton detector (500 kton-years) are simulated using the NUANCE \cite{nuance} 
neutrino generator. Honda 3D fluxes \cite{honda,honda1,honda2} for atmospheric 
neutrinos are used and the target is assumed to be a generic isoscalar one. For the \emph{perfect} case 
analyses presented in Section~\ref{ideal-chi2}, the following procedure is 
used to generate ``data'' and theory events. For ``data'' events, each event in the 100 year sample is 
oscillated individually applying the central values of the oscillation parameters given in 
Table.~\ref{osc-par-3sig}. This is then scaled down to the required number of years  
(10 years). For theory events, the 100 year sample is oscillated event by event 
by varying the parameters in their respective 3$\sigma$ ranges given in the same table. 
This method has no fluctuations. For the \emph{realistic} case, the 10 years of events are selected 
randomly from the unoscillated 100 year sample and oscillated individually with the central values 
in Table.~\ref{osc-par-3sig} to generate ``data''. The remaining 90 years of events are oscillated 
with parameters in their 3$\sigma$ ranges and scaled to 10 years to generate theory. This method thus 
takes into account the fluctuations. 

A poissonian $\chi^2$ analysis as described in \cite{hi-mu} is performed with three final state observables 
$E^{obs}_\mu,\cos\theta^{obs}_{\mu},E'^{obs}_{had}$, which are the energy and direction of 
the observed muon and the energy of the observed hadron shower. The binning scheme is shown in 
Table.~\ref{ideal-bins}. 

\begin{table}[htp]
\centering
\begin{tabular}{|c|c|c|c|}
\hline
Observable & Range & Bin width & No.of bins \\ 
\hline
& [0.1, 0.2] & 0.1 & 1 \\
& [0.2, 0.4] & 0.2 & 1 \\
& [0.4, 0.5] & 0.1 & 1 \\
& [0.5, 1.0] & 0.3 & 2 \\
& [1.0, 4.0] & 0.5 & 6 \\
& [4, 7] & 1 & 3 \\
$E^{obs}_{\mu}$ (GeV)& [7, 11] & 4 & 1 \\
(18 bins) & [11, 12.5] & 1.5 & 1 \\
& [12.5, 15] & 2.5 & 1 \\
& [15, 30] & 15 & 1 \\
\hline
$\cos\theta^{obs}_{\mu}$ (20 bins) & [-1.0, 1.0] & 0.1 & 20 \\
\hline
& [0, 2] & 1 & 2 \\
$E'^{obs}_{had}$ (GeV) & [2, 4] & 2 & 1\\
(4 bins) & [4, 15] & 11 & 1 \\
\hline
\end{tabular}
\caption{Bins of the three observables used for the analysis.}
\label{ideal-bins}
\end{table} 

Systematic uncertainties are taken into account using pull method 
\cite{hi-mu,pi6,Kameda,Ishitsuka,Gonzalez-Garcia-pull,Fogli,Huber,3D-MMD}. 
For the idealisitc case where neutrino and anti-neutrino events can be separated from each other, 
the 11 pull $\chi^2$ analysis described in \cite{hi-mu} is performed. For the realistic case when 
neutrino and anti-neutrino events cannot be separated from each other, the $\chi^2$ described by 
Eqn.~10 of \cite{lowE-atmos-dcp} is used. The parameters $\theta_{23},|\Delta{m^2}_{32}|$ and $\alpha_3$ are 
marginalised in their 3$\sigma$ ranges. The other parameters $\theta_{12},\theta_{13}$ and $\Delta{m^2}_{21}$ 
are measured precisely, so they are kept fixed in the analysis. 

In the realistic case, the effect of final state lepton energy resolution on the sensitivity to 
$\delta_{CP}$ also is studied. For this, resolutions of the form 
\cite{SK-res}:
\begin{equation}
\frac{\sigma}{E} = \frac{a\%}{\sqrt{E}} \oplus b\%~
\end{equation}
were taken, where $a=2.5$ and $b=0.5$ for electrons and $a=3$ for muons. 

\section{Results-Idealistic case}\label{ideal-chi2}
 Sensitivity studies with and without final state lepton energy resolutions 
 The results of sensitivity studies with and without pulls and energy resolutions are obtained. 
 \subsection{No pulls}
\begin{figure}[htp]
\includegraphics[width=0.45\textwidth,height=0.45\textwidth]{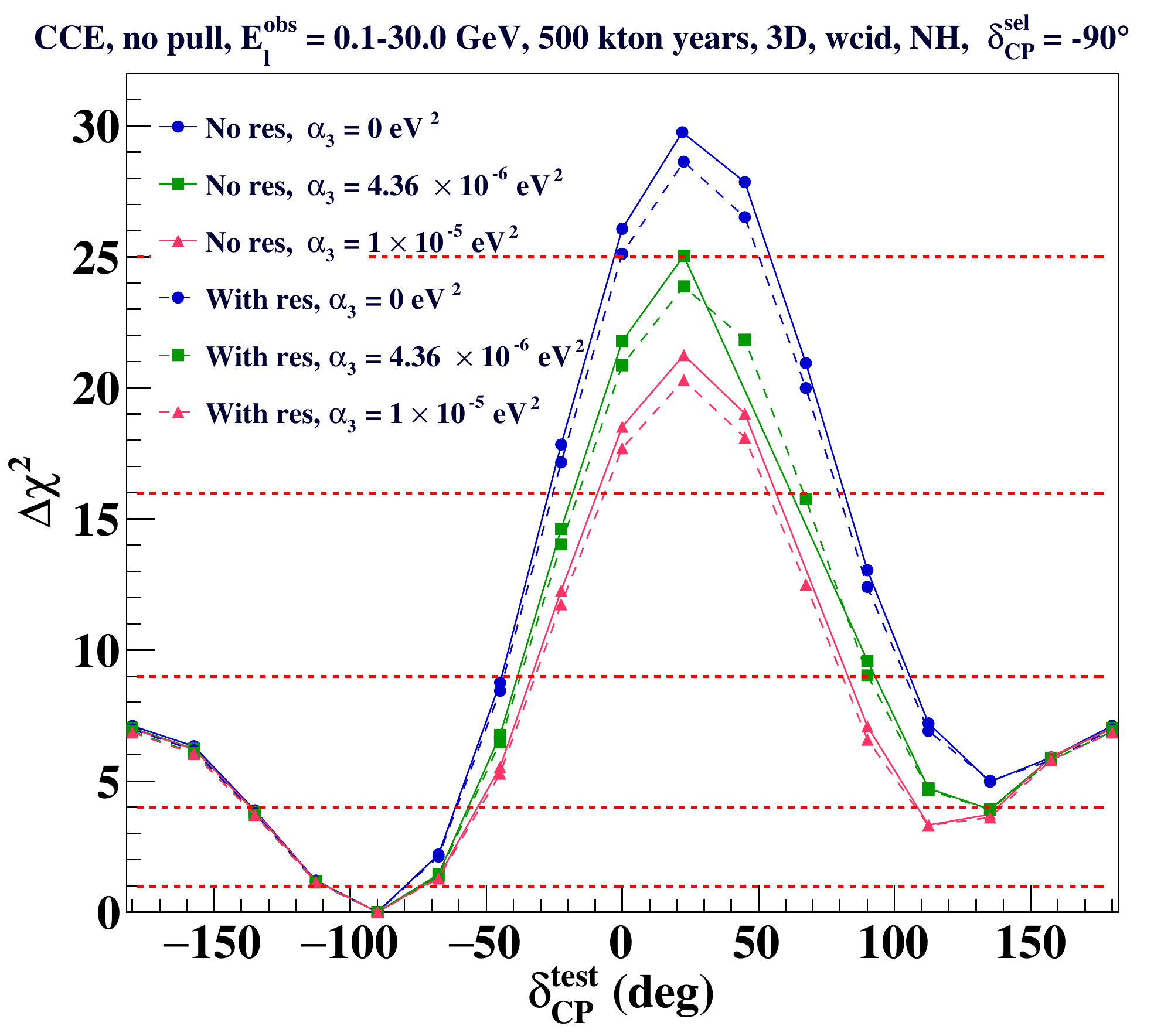}
\includegraphics[width=0.45\textwidth,height=0.45\textwidth]{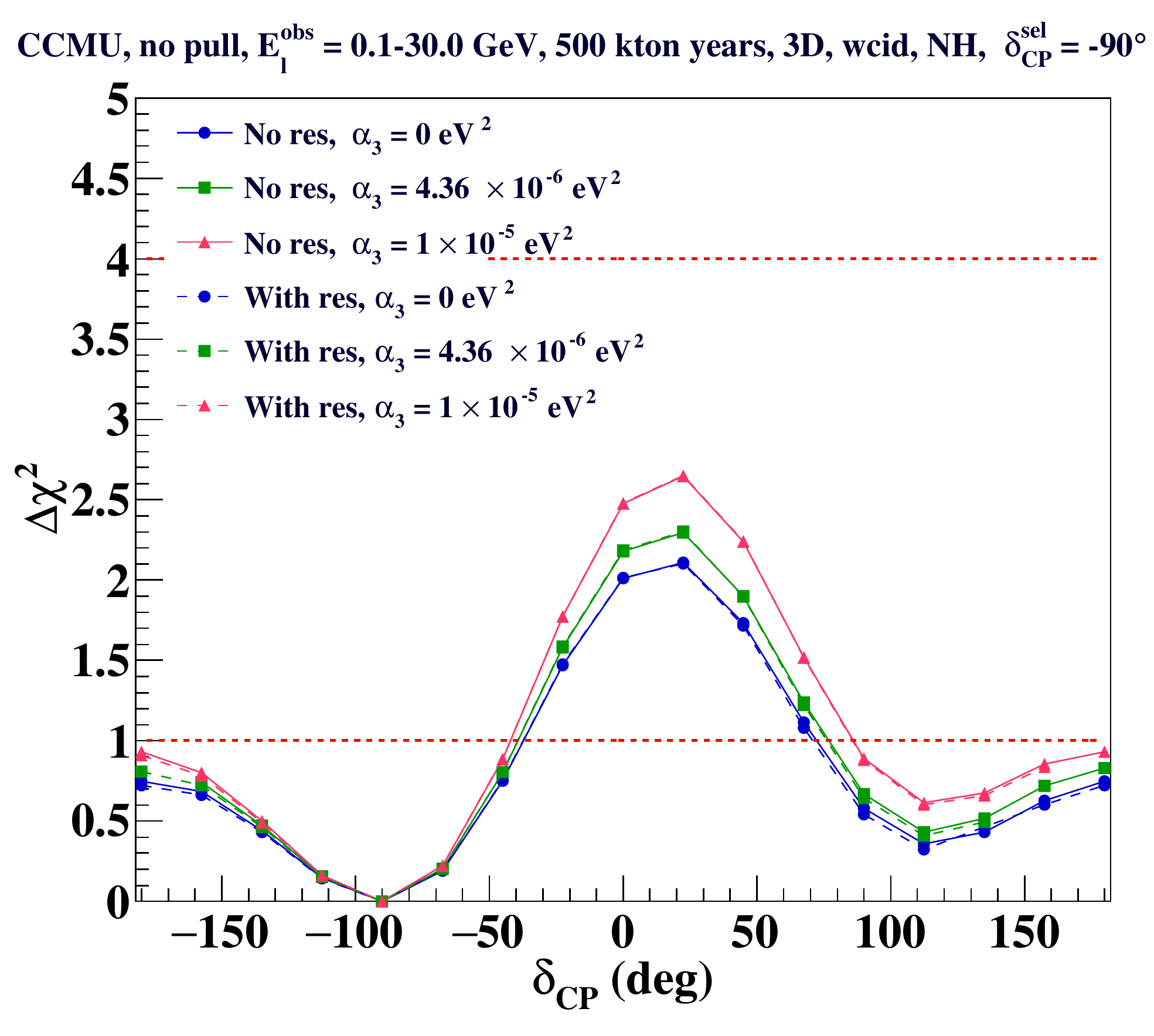}
\caption{Comparison of sensitivity $\chi^2$ with $\delta^{sel}_{CP}=-90^\circ$ and true NH for CCE (left) and 
CCMU (right) events with charge identification for the cases with and without energy resolution when no 
systematic uncertainties are present (ideal). Note that the y-axes are not identical.} 
\label{chi2-nores-nop}
\end{figure}

Fig.~\ref{chi2-nores-nop} shows the sensitivity to $\chi^2$ to the most ideal (and 
currently impractical) scenario. Several observations can be made from this figure. 
\begin{itemize}
\item The sensitivity to $\delta_{CP}$ with CCE events is much higher than that with CCMU events as expected. 
\item While the sensitivity from CCE decreases with increase in the decay parameter, the sensitivity to $\delta_{CP}$
from CCMU events is slightly enhanced in the presence of invisible decay. The effect of $\alpha_3$ is opposite on 
CCE and CCMU events. 

\item In the absence of systematic uncertainties, for a given $\alpha_3$ value, energy resolution worsens the 
sensitivity only slightly.

\item $\delta^{test}_{CP}$ values in the range $[-50^\circ,110^\circ]$ can be excluded at $3\sigma$
with CCE events alone, even in presence of invisible decay (with $\alpha_3=1\times10^{-5}$ eV$^2$). 
All values of $\delta_{CP}$ are allowed at 2$\sigma$ from CCMU events, $\delta_{CP}=\sim[-40^\circ,90^\circ]$
can be excluded at 1$\sigma$ for all three values of $\alpha_3$. 
\end{itemize}

The major contribution to the $\delta_{CP}$ sensitivity comes from the lower energy region where the effect of 
$\alpha_3$ is also high.


\subsection{Effect of pulls}\label{pull-eff-ideal}
\begin{figure}[htp]
\includegraphics[width=0.45\textwidth,height=0.45\textwidth]{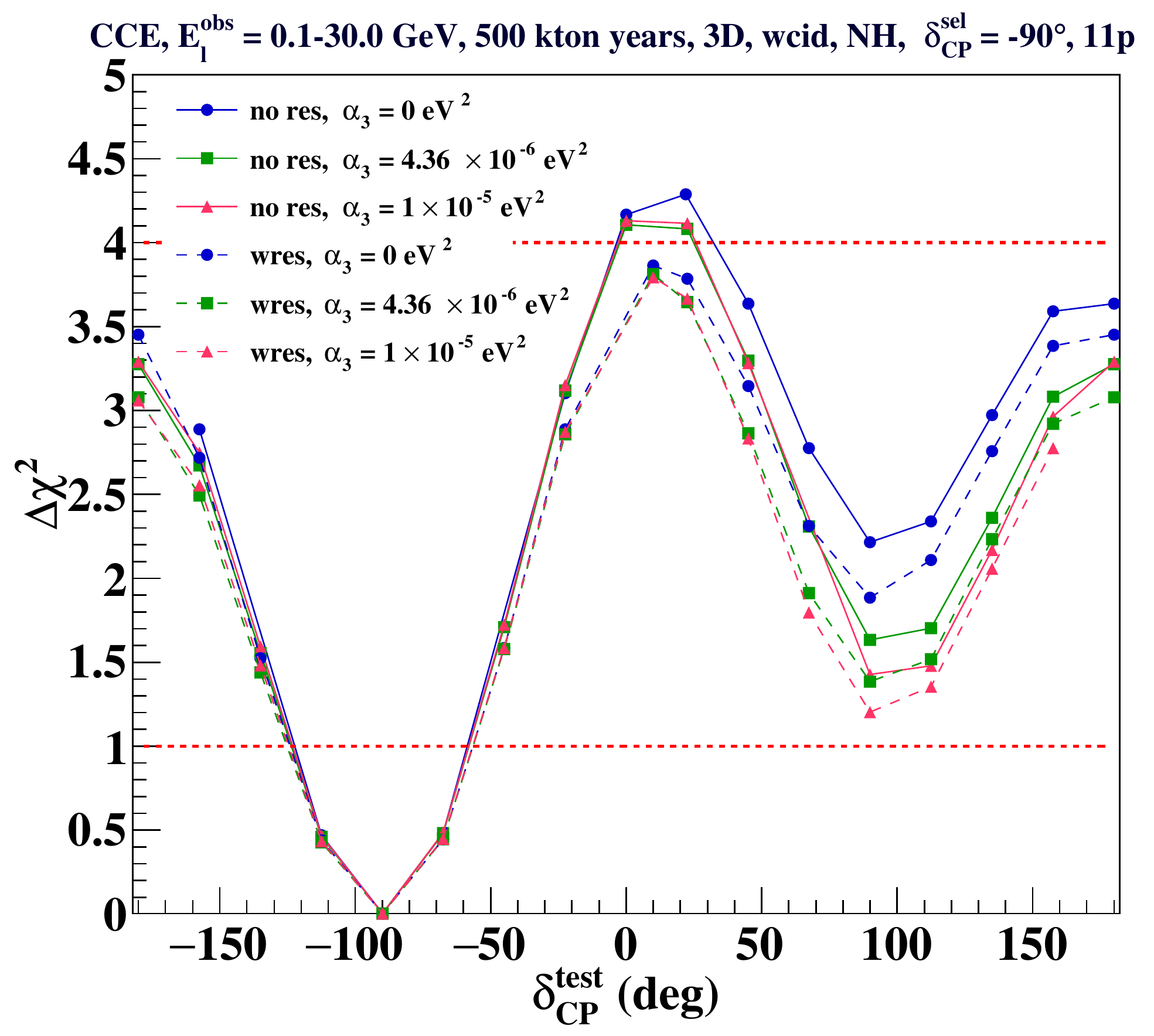}
\includegraphics[width=0.45\textwidth,height=0.45\textwidth]{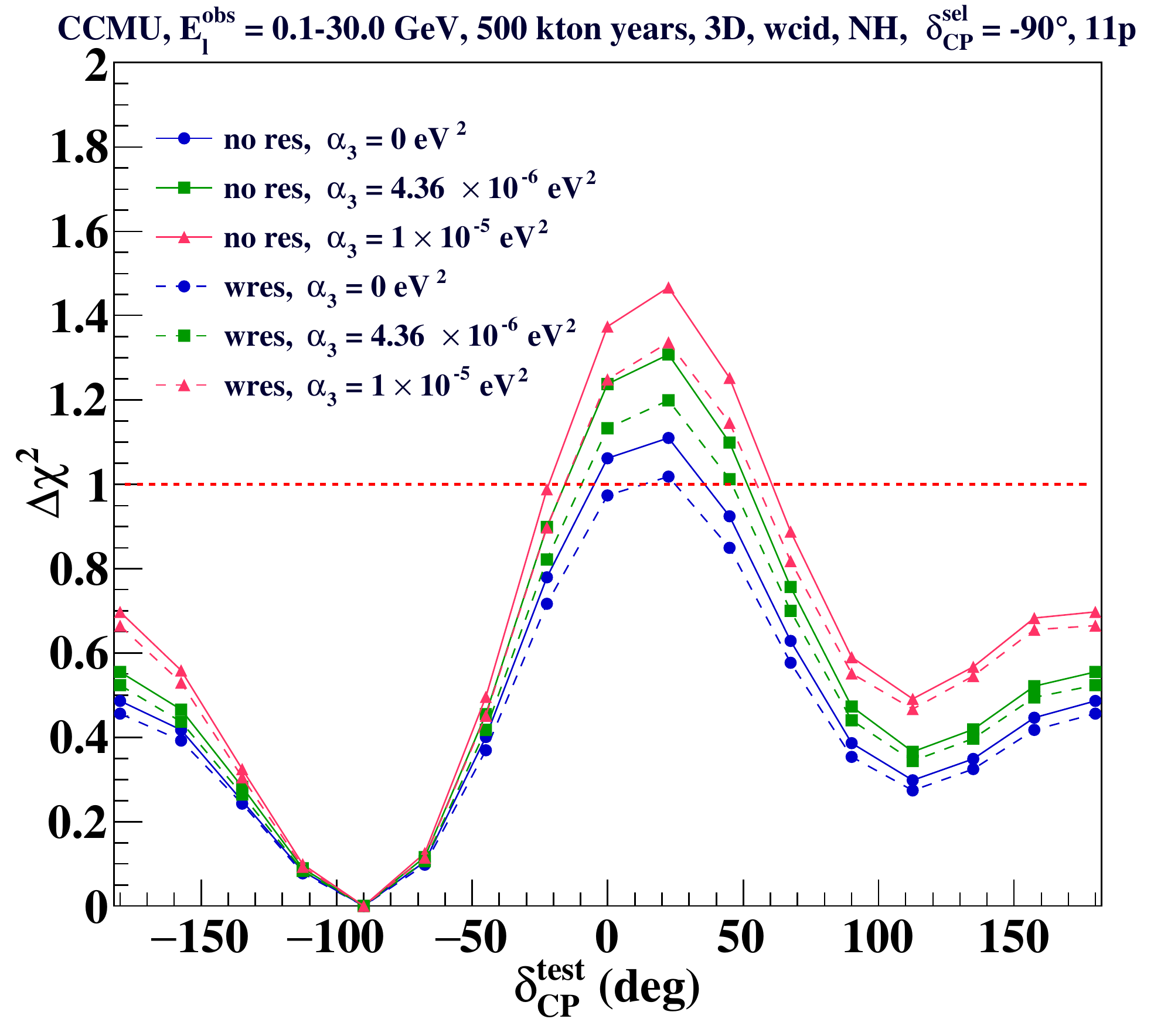}
\caption{Comparison of sensitivity $\chi^2$ with $\delta^{sel}_{CP}=-90^\circ$ and true NH for CCE and CCMU events with 
charge identification for the cases wtih and without energy resolution and with 11 pulls.} 
\label{nores-wres-11p}
\end{figure}

    When systematic unceratinties are present the sensitivity decreases significantly for both CCE and 
    CCMU events (for the idealistic CCE case this is very drasitc). For both CCE and CCMU, the sensitivities 
    with finite detector resolutions are lesser than those with perfect resolutions even in the presence of all 
    pulls. These are shown in Fig.~\ref{nores-wres-11p}. 

    \begin{itemize}
     \item In the absence of any pull, effect of $\alpha_3$ on the sensitivity was clearly visible, especially for 
     CCE events. With all 11 pull, not only does the $\delta_{CP}$ sensitivity for each value of $\alpha_3$ reduce, 
     but the distinction between the sensitivities for different $\alpha_3$ values disappear in the region 
     $\delta_{CP}=\sim[-90^\circ,-20^\circ]$. For a detector with perfect resolutions, the no-decay case will have the 
     most sensitivity even in the presence of all pulls in the region $\delta_{CP}-\sim[-20^\circ,180^\circ]$.
     There is a mildly significant separation between the no-decay and with decay cases in this $\delta_{CP}$ region.      
     The trends are similar with finite detector resolutions. 
     \item For CCMU events with 11 pulls, the effects for all 3 $\alpha_3$ values are similar until $\delta_{CP}=\sim-40^\circ$. Unlike the CCE events the separation between the sensitivities with different 
     $\alpha_3$ values can be seen well in $\delta_{CP}=\sim[-40^{\circ},180^\circ]$ in the zoomed in version. 
     The trend is similar for a finite resolution case.
     \end{itemize}
     
\begin{figure}[htp]
\includegraphics[width=0.45\textwidth,height=0.45\textwidth]{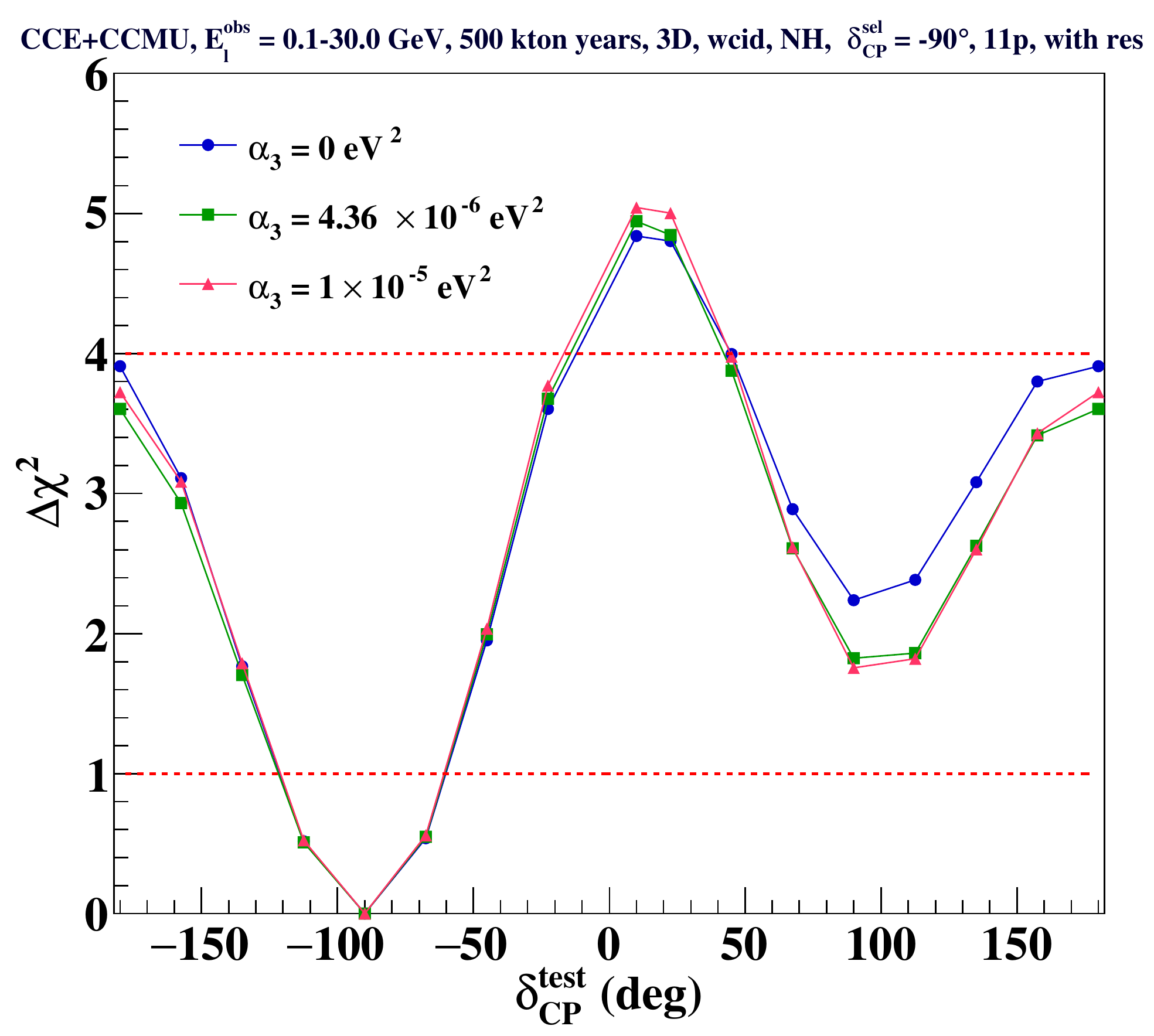}
\caption{Comparison of sensitivity $\chi^2$ with $\delta^{sel}_{CP}=-90^\circ$ and true NH for CCE and CCMU events with 
charge identification for the cases wtih and without energy resolution and with 11 pulls.} 
\label{chi2-wres-nores}
\end{figure}
When the sum is taken, the region between $[\sim-14^\circ-\sim44^\circ]$ is excluded at $2\sigma$. Also adding 
the $\chi^2$ contribution from CCMU events also restricts the region where $\alpha_3$ affects $\delta_{CP}$
sensitivity to $[\sim44^\circ-180^\circ]$. Here there is a reduction in sensitivity when $\alpha_3$ increases from
0 eV$^2$ to larger values, but there is no change in sensitivity while increasing $\alpha_3$ from $4.36\times10^{-6}$
to $1\times10^{-5}$ eV$^2$.

\begin{figure}[htp]
\includegraphics[width=0.45\textwidth,height=0.45\textwidth]{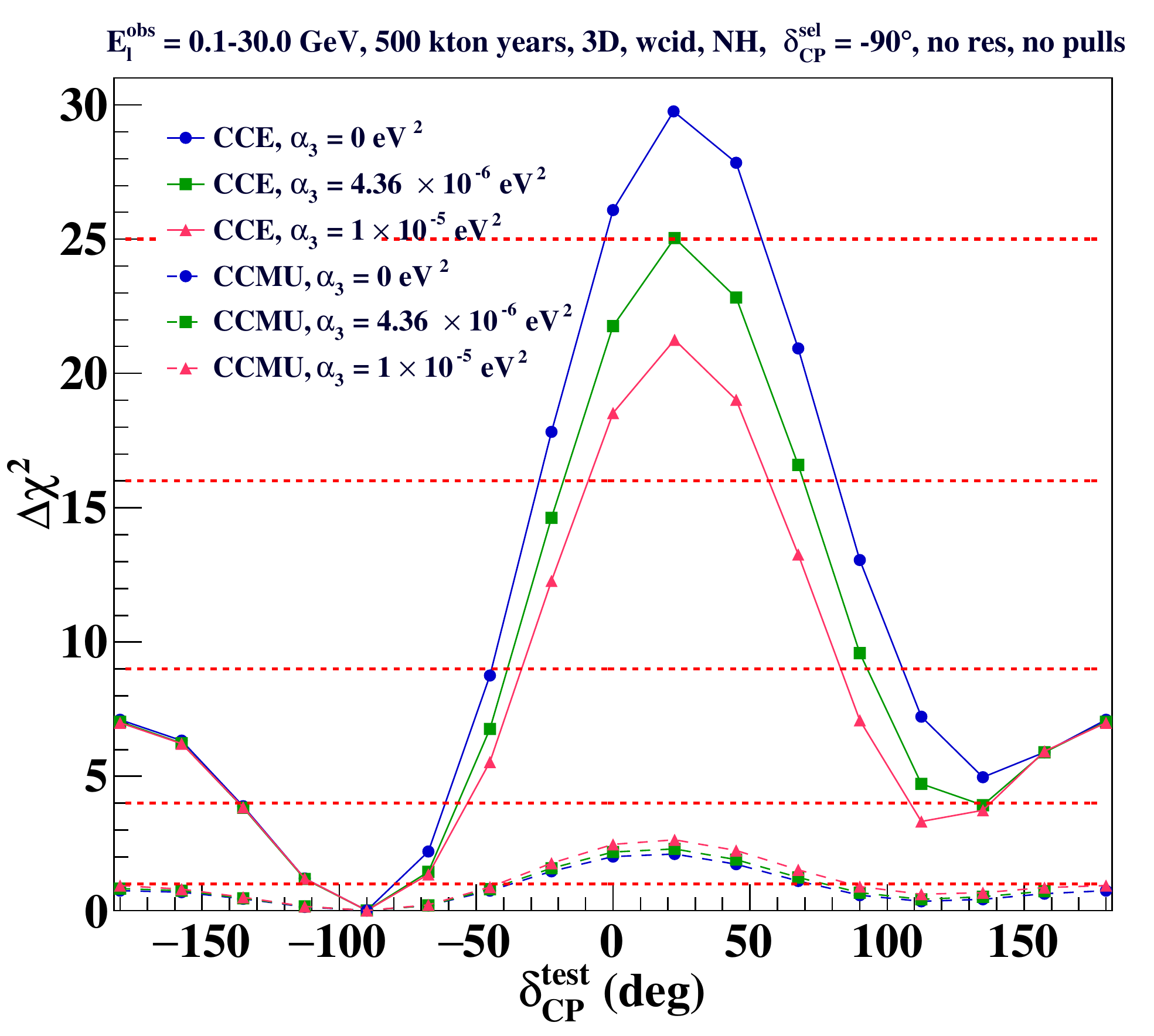}
\includegraphics[width=0.45\textwidth,height=0.45\textwidth]{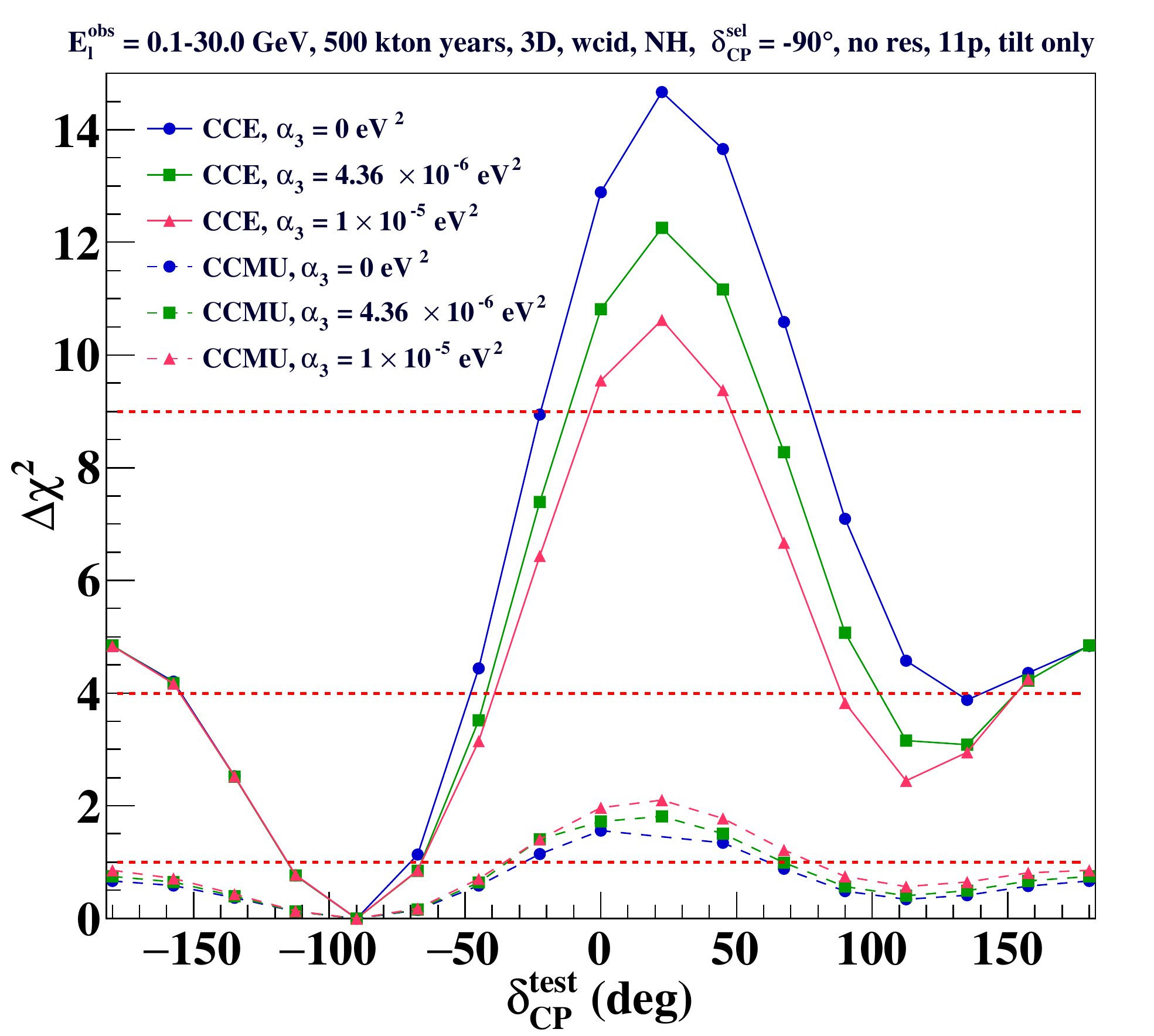}

\includegraphics[width=0.45\textwidth,height=0.45\textwidth]{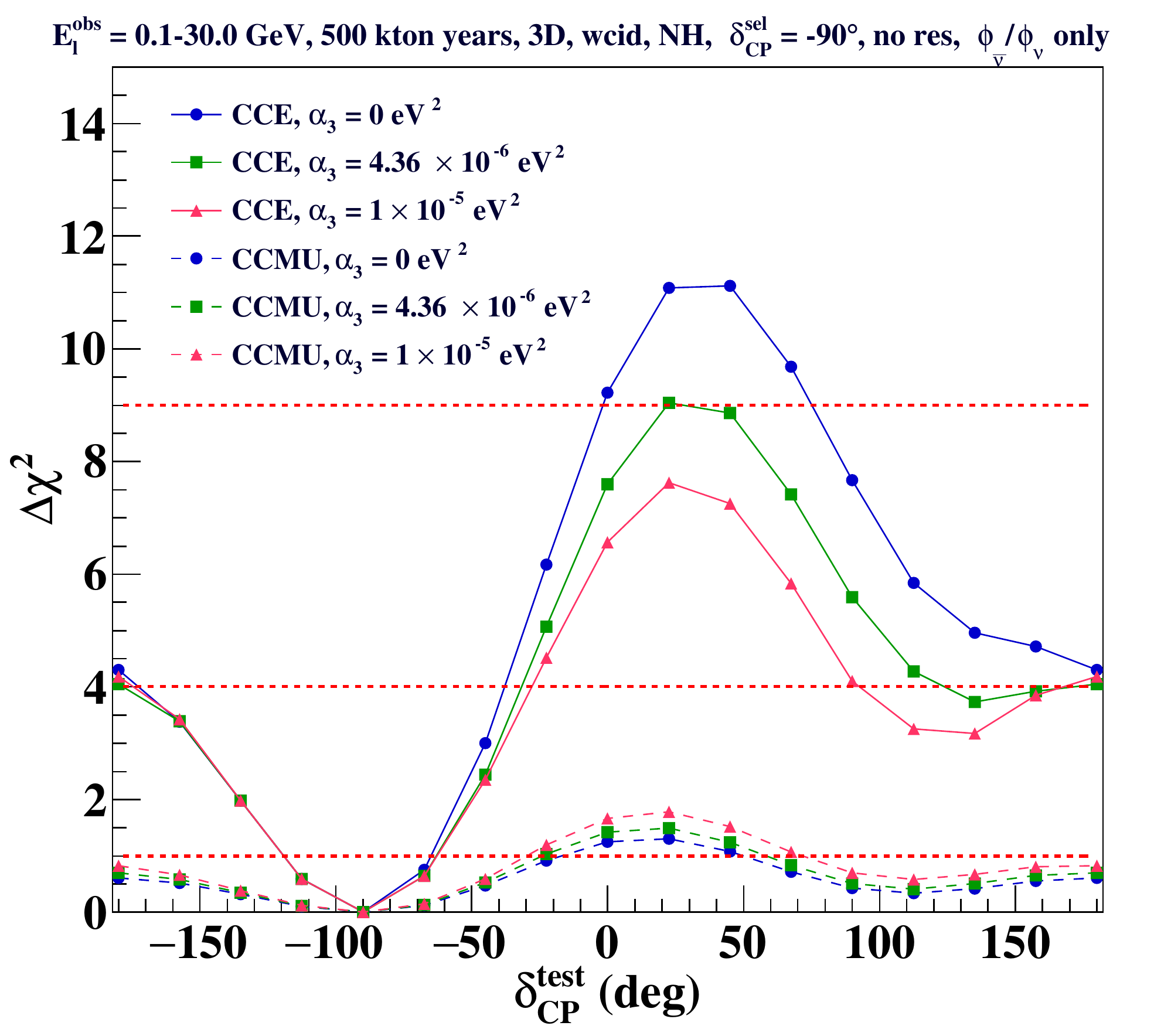}
\includegraphics[width=0.45\textwidth,height=0.45\textwidth]{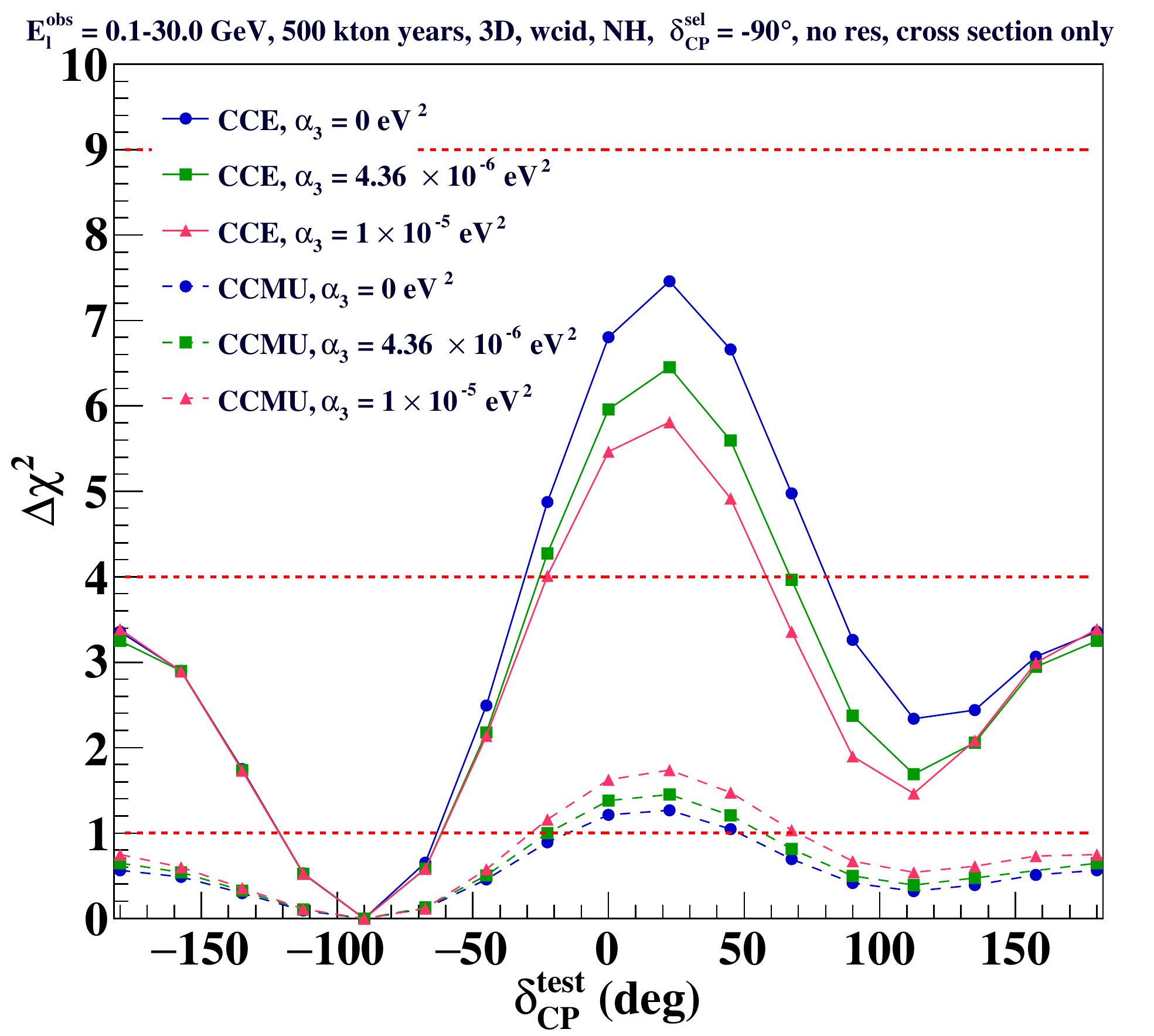}
\caption{Comparison of sensitivity $\chi^2$ with $\delta^{sel}_{CP}=-90^\circ$ and true NH for CCE and CCMU events with 
pulls switched on one by one. Y-axes are not the same.} 
\label{nores-pull-eff}
\end{figure}

      But these differences are very small and will be very difficult to separately identify in a very realistic case.
      To understand which systematic uncertainty is driving the loss of sensitivity to $\delta_{CP}$ let us look 
      at $\delta_{CP}$ $\chi^2$ for perfect resolution cases. The uncertainties - those in tilt, flux ratio and 
      cross section are switched on one each at a time. The results are shown in Fig.\ref{nores-pull-eff}. From 
      the figure it can be seen that the flux and cross section uncertainties alone can result in the reduction of 
      $\chi^2$ to about half of the no pull values for all three $\alpha_3$ values. Out of flux and cross section, the 
      cross section has more effect than the flux uncertainty on CCE events. When both these uncertainties are 
      combined we lose a significant amount of sensitivity as seen from the 11 pulls case in Fig.\ref{nores-wres-11p}.
      Hence it is important that, we measure the neutrino fluxes and cross sections precisely. 
 
\section{Results-Realisitc case}\label{real-chi2}
\begin{figure}[htp]
\includegraphics[width=0.45\textwidth,height=0.45\textwidth]{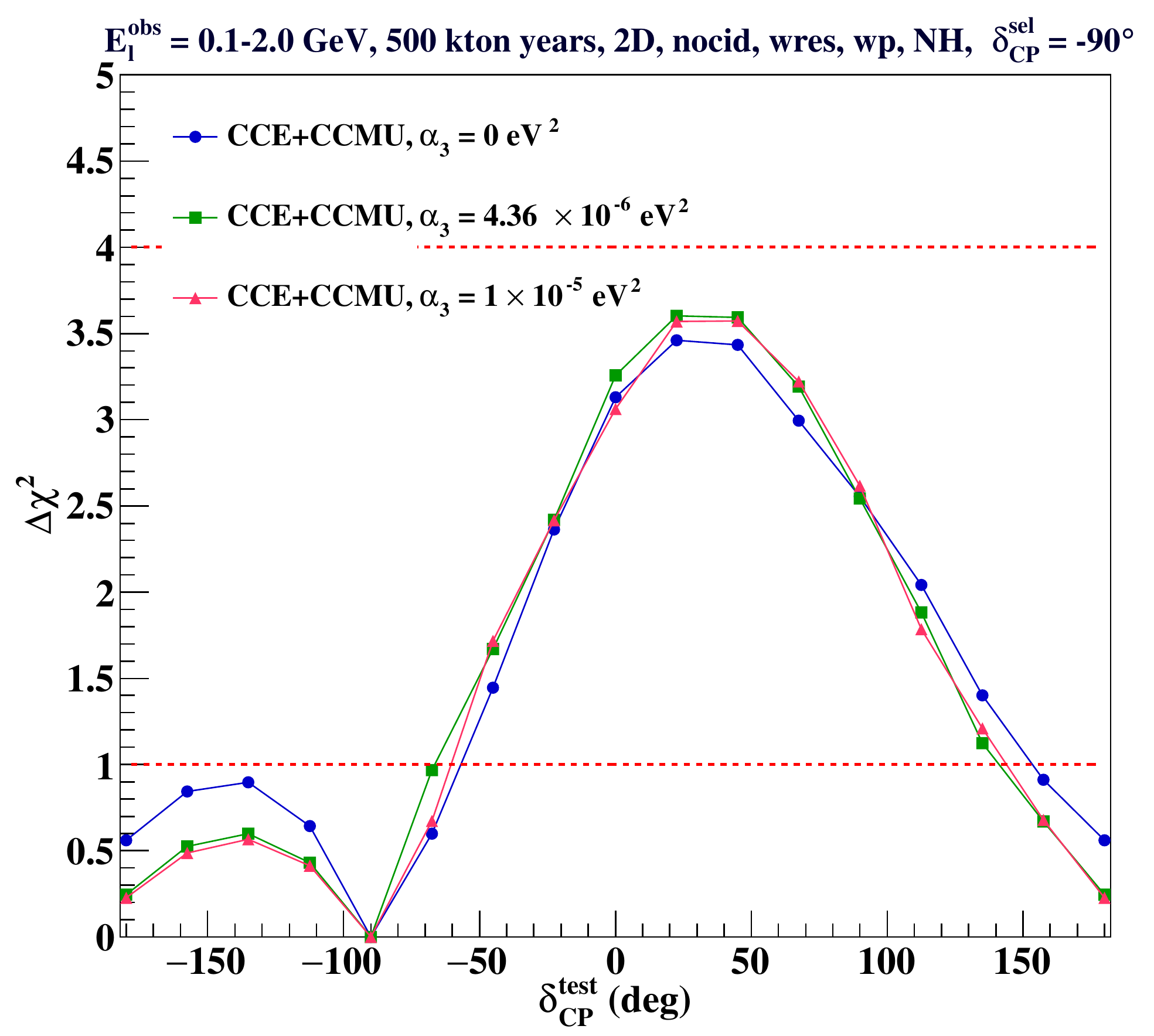}
\includegraphics[width=0.45\textwidth,height=0.45\textwidth]{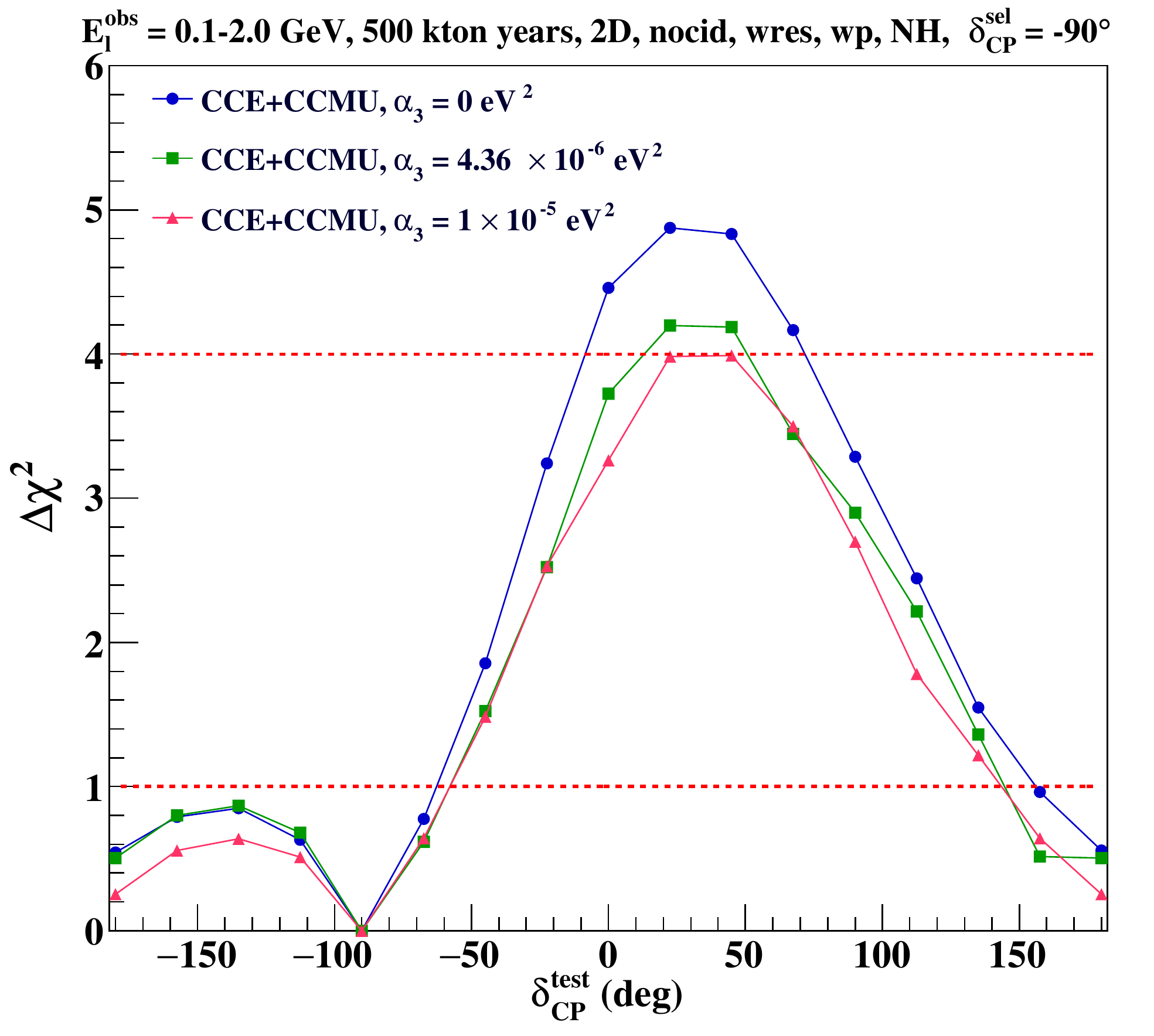}
\caption{Comparison of combined sensitivity with CCE and CCMU events in the $E^{obs}_l$ range 0.1--2.0 GeV 
with $\delta^{sel}_{CP}=-90^\circ$ and true NH for the realisitc case with (left) 5 pulls and (right) 3 pulls.
Y-axes are not the same.} 
\label{real-chi2-sum-low}
\end{figure}

   The results for the realistic cases are discussed in this section. Here the effect of fluctuations are taken into 
   account and the detector cannot separate between neutrinos and anti-neutrinos. Since there is no $\nu-\bar{\nu}$
   separation, only 5 pulls are there - those on flux (20\%), cross section (10\%), tilt (5\%), overall (5\%) and zenith angle (5\%) uncertainties. 
   For the 0.1--2.0 GeV energy range, all values of $\delta_{CP}$ are allowed at 2$\sigma$. The left panel of 
   Fig.~\ref{real-chi2-sum-low} shows this result. When all five uncertainties are present and their values are large,
   all values of $\delta_{CP}$ are allowed at 2$\sigma$ for all values of $\alpha_3$. Also all the large unceratinties 
   wash away the effect of decay and there is no way to distinguish if decay has any effect on the $\delta_{CP}$ 
   sensitivity (except in the range [-180$^\circ$,-90$^\circ$] where the sensitivity to $\delta_{CP}$ is higher for the 
   no decay case compared to the other two $\alpha_3$ values; but all of these are below 1$\sigma$ and hence are not 
   very significant).  The right panel of Fig.~\ref{real-chi2-sum-low} shows the sensitivity when there are lesser  
   and smaller uncertainties. Here, only 3 uncertainties are considered - 5\% in cross section, 5\% overall uncertainty
   and 5\% tilt. Not only do the the sensitivities increase with smaller and fewer uncertainties, but the effect of 
   $\alpha_3$ also becomes clearer between the no decay and the with decay cases. While the sensitivities of the 
   with-decay cases are similar, the reduction in sensitivity with increase of $\alpha_3$ from a no decay to with 
   decay is visible here, although it is small. For $\alpha_3=0~eV^2$ ($\alpha_3=4.36\times10^{-6}~eV^2$) the 
   $\delta_{CP}$ region $\sim$[-8$^\circ$,73$^\circ$] ($\sim$[14.5$^\circ$,51$^\circ$]) is ruled out at 2$\sigma$. All 
   values of $\delta_{CP}$ are allowed at 2$\sigma$ for $\alpha_3=1\times10^{-5}~eV^{2}$. 

   Thus, in presence of unceratinties, finite resolutions and fluctuations, the effect of 
   invisible decay of $\nu_3$ on $\delta_{CP}$ measurement is washed out. If we have to identify this effect, there 
   should be a precise measurement of fluxes and cross sections as mentioned in Section.\ref{pull-eff-ideal}.
   It can also be seen that though the contribution of CCMU events itself is very small, adding it to the CCE 
   $\chi^2$ can improve the sensitivities slightly. Since every event is valuable in low counting experiments, it 
   is worthwhile keeping these events in the analysis. 
   
  \section{Summary and conclusions}\label{summary}
   Low energy (sub GeV) atmospheric neutrino oscillations are very interesting and can help us understand the 
   neutrino oscillation parameters \cite{lowE-atmos-dcp,dune-atmos-delcp,wawu-dcp,lowE-peres-smirnov} and new physics
   scenarios like invisible neutrino decay. The effect of invisible decay of $\nu_3$, which is a new physics 
   scenario, on the measurement of $\delta_{CP}$, which is a standard neutrino oscillation parameter,
   using atmospheric neutrinos in the energy ranges 0.1--2.0 GeV and 0.1-30.0 GeV are studied for idealistic and 
   realistic cases. In the absence of 
   systematic uncertainties and with a detector having perfect resolutions the effect of $\alpha_3$ is identifiable.
   The major contribution to the sensitivity $\chi^2$ is from the energy range 0.1--2.0 GeV for both CCE and CCMU events.
   CCE events contribute more to $\delta_{CP}$ sensitivity. Sensitivity decreases (increases) with increase (decrease) 
   in $\alpha_3$ for CCE (CCMU). Presence of systematic uncertainties reduce the sensitivities drastically - flux and 
   cross section unceratinties are mainly responsible for this reduction. In the realistic case, any effect of invisible 
   decay is washed out and the sensitivity is practically the same for all values $\alpha_3$ if there are large 
   and more uncertainties. For smaller and fewer uncertainties the sensitivity improves and the effect of invisible
   decay is also discernible to a certain extend. The main unceratinties which affect the sensitivity are again those 
   in flux and cross sections. Finite detector resolutions and fluctuations also contribute to the worsening of the
   sensitivity. Hence the limitations in the detector resolution and systematic
   uncertainties can result in the non-identification of the effect of invisible decay if $\nu_3$ indeed decays in 
   nature. i.e, even if decay can affect the sensitivity to $\delta_{CP}$, with a detector without a high energy 
   resolution and uncertainties in fluxes and cross sections we will not be able to identify that effect at all.  
   This means that we need detectors with better energy resolutions and especially for atmospheric neutrinos where 
   the fluxes cannot be controlled, a precise measurement of the neutrino--anti-neutrino fluxes \cite{atmos-xenon}. 
   CCMU events get more affected by $\alpha_3$ than CCE events, especially in the very low energy bins. From the 
   oscillograms in Fig.\ref{delPmumu-delcp-a3}, this is clearly visible at energies between $\sim$ 0.1--0.2 GeV. To 
   probe these energies, a detector with a very fine energy resolution is required. Also the separation of other events 
   which can act as a background to the CCMU events in this extremely low energy range should also be possible. 
   This study is beyond the scope of this paper and has to be done in a detailed manner elsewhere. 
   In conclusion, invisble decay of $\nu_3$, if it exists in nature will have an effect on $\delta_{CP}$ measurement 
   using atmospheric neutrinos. But this can be measured perfectly only in a very idealistic scenario or atleast in a 
   case where we have good resolutions and lesser and fewer systematic uncertainties.

\section{Acknowledgement} 
   I acknowledge Prof. James Libby, Indian Institute of Technology Madras (IITM) and IITM Dept.of Physics where I started  
   doing this work. I thank Prof.D.Indumathi, The Institute of Mathematical Sciences (IMSc) for the unoscillated NUANCE 
   data and pull files. Many thanks to IMSc system administrators for the help with Nandadevi cluster on which 
   the simulations for this paper were run. I dedicate this paper to all the medical staff, doctors, nurses, health workers, 
   first responders, care givers, essential service providers and all other people who have been bravely helping others during
   Covid-19 pandemic.

  \end{document}